\documentclass[english,preprint]{aastex}
\usepackage[T1]{fontenc}
\usepackage[latin9]{inputenc}
\usepackage{amsbsy}
\usepackage{amssymb}
\usepackage{esint}
\usepackage{babel}
\begin{document}

\title{Non-Zeeman Circular Polarization of Molecular Rotational Spectral
Lines}

\author{Martin Houde$^{1,2}$, Talayeh Hezareh$^{3}$, Scott Jones$^{1}$,
and Fereshte Rajabi$^{1}$}

\affil{$^{1}$Department of Physics and Astronomy, The University of Western
Ontario, London, ON, N6A 3K7, Canada}

\affil{$^{2}$Division of Physics, Mathematics and Astronomy, California
Institute of Technology, Pasadena, CA 91125 }

\affil{$^{3}$Max-Planck-Institut f\"{u}r Radioastronomie, Auf dem H\"{u}gel
69, 53121 Bonn, Germany}
\begin{abstract}
We present measurements of circular polarization from rotational spectral
lines of molecular species in Orion KL, most notably $^{12}\mathrm{CO}\;\left(J=2\rightarrow1\right)$,
obtained at the Caltech Submillimeter Observatory with the Four-Stokes-Parameter
Spectra Line Polarimeter. We find levels of polarization of up to
1 to 2\% in general, for $^{12}\mathrm{CO}\;\left(J=2\rightarrow1\right)$
this level is comparable to that of linear polarization also measured
for that line. We present a physical model based on resonant scattering
in an attempt to explain our observations. We discuss how slight differences
in scattering amplitudes for radiation polarized parallel and perpendicular
to the ambient magnetic field, responsible for the alignment of the
scattering molecules, can lead to the observed circular polarization.
We also show that the effect is proportional to the square of the
magnitude of the plane of the sky component of the magnetic field,
and therefore opens up the possibility of measuring this parameter
from circular polarization measurements of Zeeman insensitive molecules.
\end{abstract}


\section{Introduction\label{sec:Introduction}}

Magnetic field studies in molecular clouds are most often conducted
through the detection of polarization signals, either in molecular
lines or dust continuum spectra. Except for Zeeman measurements on
suitable molecular species, which directly probe the strength of the
magnetic field (usually the line of sight component; see \citealt{Heiles1997,Crutcher1999,Brogan2001,Falgarone2008}),
all other types of observations and analyses provide indirect characterizations
of magnetic fields. Perhaps the only technique that does not rely
on polarization measurements is that of \citet{Houde2000a} and \citet{Li2008},
which is based on the comparison of line widths from coexistent molecular
ion and neutral species (see also \citealt{Houde2000b,Houde2001,Tilley2010,Tilley2011,Falceta2010}).

For a couple of decades or so measurements of dust continuum polarization,
either in emission or absorption \citep{Heiles2000,Novak2004,Li2006,Matthews2009,Dotson2010,Vaillancourt2012},
have been a particularly efficient of way of conducting systematic
studies on the interstellar medium and the effects and role of magnetic
fields in the processes leading to the formation of stars \citep{Girart2006,Attard2009}.
Although it has been known since the work of \citet{CF1953} that
polarization maps can be used to provide an estimate of the strength
of the plane of the sky component of the magnetic field, recent developments
have rendered possible refinements of this technique and the further
characterization of magnetized turbulence and its power spectrum \citep{Houde2004,Heyer2008,Hildebrand2009,Houde2009,Houde2011,Shadi2012}.

Linear Polarization of molecular lines through the so-called Goldreich-Kylafis
effect \citep{Goldreich1981}, which was first confirmed observationally
in the envelope of an evolved star some 16 years after its theoretical
prediction \citep{Glenn1997}, is now routinely detected in star-forming
regions. This effect specifies the conditions leading to an imbalance
in the population of magnetic sub-levels responsible for the $\pi$-
and $\sigma$-transitions, which are respectively polarized parallel
and perpendicular to the orientation of the plane of the sky component
of the magnetic field. Depending which sub-level populations dominate,
the detected linear polarization can be oriented in either directions.
It is in principle possible to lift this degeneracy through the observations
of several molecular transitions \citep{Cortes2005}. With the development
of array heterodyne receivers it is to be expected that this technique
will also become an important tool for mapping magnetic fields on
large scales in the interstellar medium.

The prediction and observations of circular polarization signals,
beyond the measurement of the Zeeman effect, have proven even more
challenging. But improvements are happening on the observational front.
Recent observations by \citet{Munoz2012} have successfully revealed
the detection of continuum circular polarization levels of about $1\%$
at wavelengths of 1.3 mm and $860\:\mu$m in Sgr A{*}. They attribute
the presence of such signals to a conversion of linear polarization
to circular polarization (or Faraday conversion). For molecular lines,
\citet{Cotton2011} have reported the clear detection of non-Zeeman
circular polarization of SiO maser lines in the asymptotic branch
giant star IK Tau with the VLBA. They found high levels of circular
polarization (often exceeding 10\%) that can also amount to a significant
fraction of the linear polarization simultaneously detected. They
were unable to explain their observations using alternate models for
the conversion of linear polarization to circular polarization based
on populations imbalance (e.g., \citealt{Wiebe1998,Deguchi1985}).
It thus appears that a novel physical model is needed to explain the
presence of such polarization signals in molecular lines.

This is what we endeavor to accomplish in this paper, where we present
the recent detection of circular polarization signals in rotational
spectral lines of molecular species in Orion KL, most notably $^{12}\mathrm{CO}\;\left(J=2\rightarrow1\right)$
(Sections \ref{sec:Observations} and \ref{sec:Results}). We also
introduce a physical model based on resonant scattering in an attempt
to explain our observations (Section \ref{sec:Analysis}). We follow
with a discussion in Section \ref{sec:Discussion} and end with a
short summary in Section \ref{sec:Conclusion}.

\section{Observations\label{sec:Observations}}

The observations discussed in this paper were obtained at the Caltech
Submillimeter Observatory (CSO) with the Four-Stokes-Parameter Spectral
Line Polarimeter (FSPPol; \citealt{Hezareh2010}). FSPPol allows measurements
of linear and circular polarization signals by the insertion of half-wave
(HWP) and quarter-wave (QWP) plates in the CSO telescope beam before
detection using the facility's receivers, which are sensitive to a
single state of linear polarization. Although FSPPol can operate in
the 230 GHz, 345 GHz, and 492 GHz bands with the corresponding receivers,
the observations presented in this paper were all conducted in the
230 GHz band. More details on the functioning of FSPPol can be found
in \citet{Hezareh2010}.

The circular polarization observations of $^{12}\mathrm{CO}\;\left(J=2\rightarrow1\right)$
at $230.5\:\mathrm{GHz}$ presented and discussed in this paper were
obtained on 23 November 2011 under mediocre conditions ($\tau\left(225\,\mathrm{GHz}\right)\approx0.2$)
and on 5 February 2012 under excellent skies ($\tau\left(225\,\mathrm{GHz}\right)\approx0.03-0.04$).
Both sets of observations yielded similar results; the spectrum stemming
from the February 2012 observations is shown in Figure \ref{fig:CO_cp}.
Also discussed are subsequent circular polarization measurements aimed
at the $\mathrm{HCN}\;\left(J=3\rightarrow2\right)$ rotational transition
at $265.9\:\mathrm{GHz}$, which were realized on 8 and 9 February
2012 ($\tau\left(225\,\mathrm{GHz}\right)\approx0.05$), and shown
in Figure \ref{fig:HCN_cp}. Finally, linear polarization of $^{12}\mathrm{CO}\;\left(J=2\rightarrow1\right)$
observations obtained on 12 February 2012 ($\tau\left(225\,\mathrm{GHz}\right)\approx0.13$)
are shown in Figure \ref{fig:CO_lp}. This linear polarization spectrum
is in good agreement with the previous result of \citet{Girart2004}
(see the two central panels of their Fig. 1). All observations were
pointed to the peak position of Orion KL at $\mathrm{RA\,}(\mathrm{J2000})=05^{\mathrm{h}}35^{\mathrm{m}}14\fs5$
and $\mathrm{Dec\,}(\mathrm{J2000})=-05^{\circ}22\arcmin30\farcs4$.
The telescope efficiency was measured to be $\approx60\%$ from scans
on Jupiter, while the pointing accuracy was determined to be better
than approximately $6\arcsec$.

In order to minimize instrumental effects in the detection of polarization
signals due to pointing and calibration errors, a conservative observation
method was adopted. That is, the necessary integrations for the measurement
of linear polarization (at four HWP orientations) and circular polarization
(at two QWP orientations) were kept short at one minute ON-source,
and a temperature calibration was done before each of them (see \citealt{Hezareh2010}).
The low levels of polarization at the peaks of strong lines (i.e.,
on the order of, or a few times, 0.1\% for $^{12}\mathrm{CO}$ and
$\mathrm{HCN}$) found in all these spectra are an indication of the
benefits in using this method, and can be taken as an approximation
for the level of instrument polarization present in these observations.

\section{Results\label{sec:Results}}

Our measurement of circular polarization for $^{12}\mathrm{CO}\;\left(J=2\rightarrow1\right)$
in Orion KL reveals a clear detection of polarization levels of up
to approximately 1 to 2\% across the spectral line, as can be seen
in Figure \ref{fig:CO_cp}. Shown are the Stokes $I$ spectrum, uncorrected
for telescope efficiency, and polarization levels (symbols with uncertainty,
using the scale on the right) in the top panel, while the Stokes $V$
spectrum is displayed in the bottom panel. All polarization data satisfy
$p\geq3\sigma_{p}$, where $p$ and $\sigma_{p}$ are the polarization
level and its uncertainty, respectively. These relatively low polarization
levels are still significantly higher than the instrumentation polarization
expected with FSPPol. For example, preliminary measurements of the
Zeeman effect on $\mathrm{CN}\;\left(N=2\rightarrow1\right)$ presented
in \citet{Hezareh2010} show that less than 0.2\% of Stokes $I$ is
expected to leak into Stokes $V$. This is consistent with the polarization
level detected at the peak of our $^{12}\mathrm{CO}\;\left(J=2\rightarrow1\right)$
Stokes $V$ spectrum (and $\mathrm{HCN}\;\left(J=3\rightarrow2\right)$;
see below), which should be approximately zero in view of the very
high optical depth expected for this line. Although the probability
of getting false Stokes $V$ Zeeman profiles (i.e., in the shape of
the velocity-derivative of Stokes $I$) is non-negligible with such
observations, our spectrum shows no such obvious pattern. 

Nonetheless, steps were taken to ensure that this detection is not
spurious. First, the circular polarization of $^{12}\mathrm{CO}\;\left(J=2\rightarrow1\right)$
was measured twice, once in November 2011 and again in February 2012,
under different sky conditions with similar results. Second, we performed
independent observations aimed at the $\mathrm{HCN}\;\left(J=3\rightarrow2\right)$
transition at the same position, which is also known to be strong
in Orion KL \citep{Houde2000a}%
\footnote{Although the frequency of that line is different than the design frequency
of the QWP (i.e., $\approx266\:\mathrm{GHz}$ vs. 226 GHz), it can
be shown that the only effect caused by the error in the thickness
of the QWP at a given frequency is a reduction in sensitivity to incoming
circular polarization proportional to the cosine of the thickness
error. That is, no incident linear polarization signal can be converted
into circular polarization with the FSPPol set up. %
}; the result from these observations is shown on the left panel of
Figure \ref{fig:HCN_cp}. As can be seen from the figure, there is
no detection of circular polarization up to a level of approximately
0.1\% in $\mathrm{HCN}\;\left(J=3\rightarrow2\right)$, which is an
order of magnitude less than the $^{12}\mathrm{CO}\;\left(J=2\rightarrow1\right)$
detection of Figure \ref{fig:CO_cp}. But most interestingly, we see
the presence of circular polarization at a level of approximately
$-2\%$ at the peak of a spectral feature located at $\approx-120\:\mathrm{km}\,\mathrm{s}^{-1}$,
which is mostly dominated by the $N_{K_{a}K_{c}}=12_{1,12}\rightarrow11_{1,11}$
transitions of $\mathrm{HNCO}$ and $\mathrm{HN}^{13}\mathrm{CO}$
in the image band at 262.8 GHz; a close-up of this spectrum is shown
on the right panel of the same figure. The fact that simultaneous
measurements on two spectral features (i.e., $\mathrm{HCN}$ and $\mathrm{HNCO}/\mathrm{HN}^{13}\mathrm{CO}$)
yield none and a detection is strong evidence that our observations
of $^{12}\mathrm{CO}\;\left(J=2\rightarrow1\right)$ are not spurious
but result from true circular polarization signals present in the
spectral line. In the next section we present a physical model that
seeks to explain these observations.

\section{Analysis\label{sec:Analysis}}

We consider a molecule immersed in a medium that harbors a magnetic
field, which provides spatial alignment. We further assume the existence
of incident radiation close to, or at, a resonant frequency of the
molecule and in a state of linear polarization at some angle relative
to the orientation of the magnetic field. We can imagine a situation
where a population of a given molecular species in a background medium
emits radiation that is linearly polarized, for example, through the
Goldreich-Kylafis effect, which is then incident on similar molecules
located in the foreground where the orientation of the magnetic field
has changed. We will investigate the conditions necessary to transform
linear polarization into circular polarization. Such a situation has
been previously considered in the literature (e.g., \citealt{Deguchi1985}),
but the mechanism for the generation of circular polarization presented
below is different.

\subsection{Conversion of Linear to Circular Polarization\label{sub:simple}}

We denote this incident background radiation state by $\left|\psi_{0}\right\rangle $,
which is linearly polarized at an angle $\theta$ with the foreground
magnetic field, which we assume located in the plane normal to the
direction of propagation, for simplicity (this assumption will be
dropped later on). This radiation state can be decomposed as follows

\begin{equation}
\left|\psi_{0}\right\rangle =\alpha_{0}\left|n_{\Vert}\right\rangle +\beta_{0}\left|n_{\bot}\right\rangle ,\label{eq:psi0}
\end{equation}

\noindent where $\alpha_{0}=\cos\left(\theta\right)$ and $\beta_{0}=\sin\left(\theta\right)$,
while $\left|n_{\Vert}\right\rangle $ and $\left|n_{\bot}\right\rangle $
are $n$-photon states linearly polarized parallel and perpendicular
to the foreground magnetic field, respectively, propagating toward
the observer. These states are orthogonal to one another and normalized.
The basic idea is to determine whether these two states scatter differently
off a foreground molecule in such a manner that a small relative phase
shift $\phi^{\prime}$ is introduced between them. The scattered radiation
state would then become (up to a global phase term)

\begin{eqnarray}
\left|\psi^{\prime}\right\rangle  & \simeq & \alpha_{0}\left(1+i\phi^{\prime}\right)\left|n_{\Vert}\right\rangle +\beta_{0}\left|n_{\bot}\right\rangle \nonumber \\
 & \simeq & \alpha_{0}e^{i\phi^{\prime}}\left|n_{\Vert}\right\rangle +\beta_{0}\left|n_{\bot}\right\rangle ,\label{eq:psi'}
\end{eqnarray}

\noindent where it was assumed that the separate phase shifts for
both states are much less than unity (therefore $\phi^{\prime}\ll1$
follows for the last equation). It is expected, however, that a very
large number of scattering $N$ will occur for a given radiation state
as it propagates within a molecular cloud. Each scattering will contribute
a relative phase shift $\phi^{\prime}$ and the final radiation state
becomes

\begin{equation}
\left|\psi\right\rangle \simeq\alpha_{0}e^{i\phi}\left|n_{\Vert}\right\rangle +\beta_{0}\left|n_{\bot}\right\rangle \label{eq:psi}
\end{equation}

\noindent with $\phi=N\phi^{\prime}$.

We can also define $n$-photon circular polarization states with

\begin{equation}
\left|n_{\pm}\right\rangle =\frac{1}{\sqrt{2}}\left(\left|n_{\Vert}\right\rangle \pm i\left|n_{\bot}\right\rangle \right).\label{eq:n+-}
\end{equation}

\noindent It is easy to show that the level of circular polarization
(Stokes) $v$ for the final state is

\begin{eqnarray}
v & = & \left\Vert \left\langle n_{+}\right|\left.\psi\right\rangle \right\Vert ^{2}-\left\Vert \left\langle n_{-}\right|\left.\psi\right\rangle \right\Vert ^{2}\nonumber \\
 & = & -2\alpha_{0}\beta_{0}\sin\left(\phi\right).\label{eq:V}
\end{eqnarray}

We can also define the complete state of linear polarization by introducing
complementary states of linear polarization oriented at $\pm45^{\circ}$
from the magnetic field direction

\begin{equation}
\left|n_{\pm45}\right\rangle =\frac{1}{\sqrt{2}}\left(\left|n_{\Vert}\right\rangle \pm\left|n_{\bot}\right\rangle \right)\label{eq:n45}
\end{equation}

\noindent and the (Stokes) parameters $q$ and $u$ with

\begin{eqnarray}
q & = & \left\Vert \left\langle n_{\Vert}\right|\left.\psi\right\rangle \right\Vert ^{2}-\left\Vert \left\langle n_{\bot}\right|\left.\psi\right\rangle \right\Vert ^{2}\nonumber \\
 & = & \alpha_{0}^{2}-\beta_{0}^{2}\label{eq:Q}\\
u & = & \left\Vert \left\langle n_{+45}\right|\left.\psi\right\rangle \right\Vert ^{2}-\left\Vert \left\langle n_{-45}\right|\left.\psi\right\rangle \right\Vert ^{2}\nonumber \\
 & = & 2\alpha_{0}\beta_{0}\cos\left(\phi\right).\label{eq:U}
\end{eqnarray}

\noindent These can be compared with corresponding parameters for
the incident radiation state

\begin{eqnarray}
q_{0} & = & \alpha_{0}^{2}-\beta_{0}^{2}\label{eq:Q0}\\
u_{0} & = & 2\alpha_{0}\beta_{0}\label{eq:U0}\\
v_{0} & = & 0\label{eq:V0}
\end{eqnarray}

\noindent to find that, for the chosen system of reference, linear
polarization is being converted from $u_{0}$ to circular polarization
$v$, and that the total amount of polarization is conserved. 

The previous example considered the case of strictly forward scattering
where the radiation states before and after scattering were composed
of the same $\left|n_{\Vert}\right\rangle $ and $\left|n_{\bot}\right\rangle $
states. We can generalize the analysis by considering a number of
initial states $\left|\psi_{j}\right\rangle $ propagating in different
directions and/or containing different numbers of photons (but of
the same frequency) such that 

\begin{equation}
\left|\psi_{j}\right\rangle =\alpha_{j}\left|n_{j,\Vert}\right\rangle +\beta_{j}\left|n_{j,\bot}\right\rangle .\label{eq:psi_j}
\end{equation}

\noindent These states could all potentially scatter in the direction
of the observer into $\left|n_{\Vert}\right\rangle $ and $\left|n_{\bot}\right\rangle $.
The incident radiation state then becomes

\begin{equation}
\left|\psi\right\rangle =\sum_{j}\left(\alpha_{j}\left|n_{j,\Vert}\right\rangle +\beta_{j}\left|n_{j,\bot}\right\rangle \right),\label{eq:total_state}
\end{equation}

\noindent with $\sum_{j}\left(\alpha_{j}^{2}+\beta_{j}^{2}\right)=1$.
The scattered state, ascribed to $j=0$ (i.e., $\left|n_{0,\Vert}\right\rangle =\left|n_{\Vert}\right\rangle $
and $\left|n_{0,\bot}\right\rangle =\left|n_{\bot}\right\rangle $)
retains the same form as equation (\ref{eq:psi'}) (after normalization)

\begin{equation}
\left|\psi^{\prime}\right\rangle \simeq\alpha^{\prime}e^{i\phi^{\prime}}\left|n_{\Vert}\right\rangle +\beta^{\prime}\left|n_{\bot}\right\rangle \label{eq:psi'_again}
\end{equation}

\noindent when 

\begin{eqnarray}
\alpha^{\prime} & = & BC\alpha_{0}\label{eq:alpha'}\\
\beta^{\prime} & = & C\beta_{0}\label{eq:beta'}\\
\phi^{\prime} & = & \frac{1}{D}\left[\mathrm{Re}\left\{ \sum_{j}\phi_{j,\Vert}^{\prime}\frac{\alpha_{j}}{\alpha_{0}}\right\} \left(1-\mathrm{Im}\left\{ \sum_{k}\phi_{k,\bot}^{\prime}\frac{\beta_{k}}{\beta_{0}}\right\} \right)\right.\nonumber \\
 &  & \left.-\mathrm{Re}\left\{ \sum_{j}\phi_{k,\bot}^{\prime}\frac{\beta_{j}}{\beta_{0}}\right\} \left(1-\mathrm{Im}\left\{ \sum_{k}\phi_{k,\Vert}^{\prime}\frac{\alpha_{j}}{\alpha_{0}}\right\} \right)\right],\label{eq:gamma'}
\end{eqnarray}

\noindent where 

\begin{eqnarray}
B & = & \frac{1}{\left\Vert 1+i\sum_{k}\phi_{k,\bot}^{\prime}\frac{\beta_{k}}{\beta_{0}}\right\Vert ^{2}}\left[1-\sum_{j}\left(\mathrm{Im}\left\{ \phi_{j,\Vert}^{\prime}\frac{\alpha_{j}}{\alpha_{0}}\right\} +\mathrm{Im}\left\{ \phi_{j,\bot}^{\prime}\frac{\beta_{j}}{\beta_{0}}\right\} \right)\right.\nonumber \\
 &  & \left.+\sum_{j,k}\left(\mathrm{Re}\left\{ \phi_{j,\Vert}^{\prime}\frac{\alpha_{j}}{\alpha_{0}}\right\} \mathrm{Re}\left\{ \phi_{k,\bot}^{\prime}\frac{\beta_{k}}{\beta_{0}}\right\} +\mathrm{Im}\left\{ \phi_{j,\Vert}^{\prime}\frac{\alpha_{j}}{\alpha_{0}}\right\} \mathrm{Im}\left\{ \phi_{k,\bot}^{\prime}\frac{\beta_{k}}{\beta_{0}}\right\} \right)\right]\label{eq:B}\\
C & = & \frac{1+i\sum_{k}\phi_{k,\bot}^{\prime}\frac{\beta_{k}}{\beta_{0}}}{\sqrt{\alpha_{0}^{2}\left\Vert 1+i\sum_{k}\phi_{k,\Vert}^{\prime}\frac{\alpha_{k}}{\alpha_{0}}\right\Vert ^{2}+\beta_{0}^{2}\left\Vert 1+i\sum_{k}\phi_{k,\bot}^{\prime}\frac{\beta_{k}}{\beta_{0}}\right\Vert ^{2}}}\label{eq:C}\\
D & = & B\left\Vert 1+i\sum_{k}\phi_{k,\bot}^{\prime}\frac{\beta_{k}}{\beta_{0}}\right\Vert ^{2}\label{eq:D}
\end{eqnarray}

\noindent and $\mathrm{Re}\left\{ \cdots\right\} $ and $\mathrm{Im}\left\{ \cdots\right\} $
stand for the real and imaginary parts, respectively. In equations
(\ref{eq:alpha'}) to (\ref{eq:D}) $\phi_{j,\Vert}^{\prime}$ and
$\phi_{j,\bot}^{\prime}$ are the scattering amplitudes for the different
polarization states, while $\alpha_{j}$ and $\beta_{j}$ are assumed
complex in general (except for $\alpha_{0}$ and $\beta_{0}$, which
are chosen to be real). The scattering amplitudes $\phi_{j,\Vert}^{\prime}$
and $\phi_{j,\bot}^{\prime}$ will contain geometrical factors that
will account for the different incidence angles. It can be verified
that the earlier forward scattering case is recovered (up to a global
phase term) when $j=k=0$ is the only possibility, the imaginary components
are zero, and $\phi_{\Vert}^{\prime},\phi_{\bot}^{\prime}\ll1$. The
form of equation (\ref{eq:psi}) and the others that follow are therefore
still adequate for the more general case.

\subsection{Dielectric Susceptibility\label{sub:Molecular-Polarizability}}

The approach taken in the previous section may appear counter-intuitive
with respect to more common types of analyses encountered when dealing
with molecular polarizability. More precisely, it could seem more
natural to investigate the relative phase shift between the scattered
and incident radiations, due to the interaction with a molecule, through
calculations of the induced molecular dielectric susceptibility \citep{Grynberg2010,Cohen1988,Cohen1977}.
We will show here that this approach cannot account for the effect
we will discuss in this paper.

For a detailed and more realistic treatment of radiation/molecule
interactions involving two molecular energy levels it is often preferable
to use the density operator (or density matrix) $\hat{\sigma}$ to
perform the analysis. Under this formalism it is found that the $j$th
component of the mean electric dipole moment $\hat{d}_{j}$ can be
evaluated with

\begin{eqnarray}
\left\langle \hat{d}_{j}\right\rangle  & = & \mathrm{Tr}\left\{ \hat{\sigma}\hat{d}_{j}\right\} \nonumber \\
 & = & \sum_{a,b}\sigma_{ba}d_{j,ab},\label{eq:mean_d}
\end{eqnarray}

\noindent where $\mathrm{Tr}\left\{ \cdots\right\} $ denotes the
trace and $\sigma_{ba}=\left\langle b\right|\hat{\sigma}\left|a\right\rangle $
is a matrix element, with $\left|b\right\rangle $ the quantum state
for level $b$, etc. With the knowledge of $\left\langle \hat{d}_{j}\right\rangle $
one can determine the induced molecular polarization component $P_{j}=\left\langle \hat{d}_{j}\right\rangle /V$,
with $V$ the volume under consideration, and the dielectric susceptibility
$\chi_{j}$ through

\begin{equation}
P_{j}=\chi_{j}E_{0,j},\label{eq:suscep}
\end{equation}

\noindent with $\mathbf{E}_{0}$ the incident electric field. 

By performing these calculations for each linear polarization component,
one can determine the relative phase shift induced between the two
scattered electric field components from the difference in the corresponding
dielectric susceptibilities. For example, if we respectively have
$\chi_{\Vert}$ and $\chi_{\bot}$ for the susceptibilities parallel
and perpendicular to the orientation of the magnetic field responsible
for the alignment of the molecule, then the relative phase shift $\varphi$
is given by

\begin{equation}
\varphi\simeq\left(\chi_{\Vert}-\chi_{\bot}\right)\frac{\omega}{2c}\Delta L,\label{eq:phase_shift}
\end{equation}

\noindent with $\omega$, $c$, and $\Delta L$ the frequency of radiation,
the speed of light, and the propagation path length after scattering,
respectively. Such calculations, when applied to the case treated
in this paper, reveal that $\varphi$ is several orders of magnitude
too small to explain our observations. 

It is important to note that because selection rules for electric
dipole transitions specify that $d_{j,ab}=0$ when $a=b$, the only
components of the density matrix involved in these calculations are
off-diagonal elements, i.e., $\sigma_{ba}$ for $b\neq a$ (see eq.
{[}\ref{eq:mean_d}{]}). As we will see in the following section,
the resonant scattering process we study is of the second order in
the electric field and only appears in the diagonal elements of the
density matrix. More precisely, if an expansion of the density matrix
in terms of powers of the interaction Hamiltonian $\hat{H}_{\mathrm{I}}$
is used \citep{Grynberg2010}, then it can be shown that the second
order diagonal terms are

\begin{eqnarray}
\sigma_{aa}^{\left(2\right)} & = & -\frac{1}{\hbar^{2}}\sum_{b\neq a}\left(\sigma_{aa}^{\left(0\right)}-\sigma_{bb}^{\left(0\right)}\right)\int_{t_{0}}^{t}e^{-\Gamma_{a}\left(t-t^{\prime}\right)}\nonumber \\
 &  & \left[\left\langle a\right|H_{\mathrm{I}}\left(t^{\prime}\right)\left|b\right\rangle \int_{t_{0}}^{t^{\prime}}\left\langle b\right|H_{\mathrm{I}}\left(t^{\prime\prime}\right)\left|a\right\rangle e^{-\left(i\omega_{ba}+\gamma_{ba}\right)\left(t^{\prime}-t^{\prime\prime}\right)}dt^{\prime\prime}\right.\nonumber \\
 &  & \left.+\left\langle b\right|H_{\mathrm{I}}\left(t^{\prime}\right)\left|a\right\rangle \int_{t_{0}}^{t^{\prime}}\left\langle a\right|H_{\mathrm{I}}\left(t^{\prime\prime}\right)\left|b\right\rangle e^{-\left(i\omega_{ba}+\gamma_{ba}\right)\left(t^{\prime}-t^{\prime\prime}\right)}dt^{\prime\prime}\right]dt^{\prime}.\label{eq:sigma_aa^2}
\end{eqnarray}

\noindent In equation (\ref{eq:sigma_aa^2}) $\Gamma_{a}$ is the
relaxation rate of level $a$, $\omega_{ba}$ and $\gamma_{ba}$ are,
respectively, the frequency and relaxation coefficient for a transition
between states $\left|b\right\rangle $ and $\left|a\right\rangle $,
and $t-t_{0}$ is the duration of the interaction. Since the interaction
Hamiltonian involves the electric dipole moment (i.e., $\hat{H}_{\mathrm{I}}=-\hat{\mathbf{d}}\cdot\mathbf{E}$),
it brings a scattering of radiation where the molecule initially in
state $\left|a\right\rangle $ is momentarily excited to (the virtual)
state $\left|b\right\rangle $ before settling back to $\left|a\right\rangle $.
As was mentioned earlier, this process cannot be captured in calculations
involving dielectric susceptibilities described through equations
(\ref{eq:mean_d})-(\ref{eq:phase_shift}).

\subsection{Resonant Scattering\label{sub:Resonant-Scattering}}

We return to the case treated in Section \ref{sub:simple} of the
interaction between an incident radiation state and a single molecule,
which for simplicity we assume to be linear (like $^{12}\mathrm{CO}$).
We concentrate on two pairs of photon states of the type $\left|n_{\Vert}\right\rangle $
and $\left|n_{\bot}\right\rangle $, and $\left|n_{\Vert}^{\prime}\right\rangle $
and $\left|n_{\bot}^{\prime}\right\rangle $ for the incident and
scattered radiation, respectively. These states contain $n$ and $n^{\prime}$
photons, while in general $n\neq n^{\prime}$. It is the $n^{\prime}$-photon
states $\left|n_{\Vert}^{\prime}\right\rangle $ and $\left|n_{\bot}^{\prime}\right\rangle $
that are eventually detected by our system to measure polarization. 

Since we seek a process that imparts a relative phase shift between
such pairs of states, it should be clear that the absorption of a
photon followed by a spontaneous emission could not lead to the desired
effect. Such a process would randomize any relative phase difference
between the two linear polarization states at the emission stage.
We must therefore move to a higher (second) order mode of interaction.
As was mentioned earlier, the best candidate for this is the resonant
scattering process where an incident photon is absorbed into a virtual,
excited state of the molecule and then re-emitted into a scattered
radiation state. 

We now denote the initial and final molecule-radiation states as 

\begin{eqnarray}
\left|i\right\rangle  & = & \left|a\right\rangle \otimes\left|\psi_{j}\right\rangle \label{eq:initial}\\
\left|f\right\rangle  & = & \left|a^{\prime}\right\rangle \otimes\left|\psi^{\prime}\right\rangle ,\label{eq:final}
\end{eqnarray}

\noindent with $\left|a\right\rangle $ and $\left|a^{\prime}\right\rangle $
the initial and final molecular states, respectively at sufficiently
long times before and after the interaction, while $\left|\psi_{j}\right\rangle $
and $\left|\psi^{\prime}\right\rangle $ are given by

\begin{eqnarray}
\left|\psi_{j}\right\rangle  & = & \alpha_{j}\left|n_{\Vert,j}\right\rangle +\beta_{j}\left|n_{\bot,j}\right\rangle \label{eq:psi_j_again}\\
\left|\psi^{\prime}\right\rangle  & = & \alpha^{\prime}e^{i\phi^{\prime}}\left|n_{\Vert}^{\prime}\right\rangle +\beta^{\prime}\left|n_{\bot}^{\prime}\right\rangle .\label{eq:psi_prime}
\end{eqnarray}

\noindent In view of our earlier discussion we have already included
the phase factor $e^{i\phi^{\prime}}$ in equation (\ref{eq:psi_prime})
(see eq. {[}\ref{eq:psi'_again}{]}). The scattering amplitude resulting
from the interaction with a single molecule can then be determined
through (using MKS units; \citealt{Grynberg2010,Cohen1988})

\begin{equation}
S_{if,\ell}=-i\frac{T}{L^{3}}\frac{\sqrt{nn^{\prime}}}{2\epsilon_{0}\hbar\sqrt{\omega\omega^{\prime}}}\sum_{b}\frac{\omega_{ba^{\prime}}\omega_{ba}\left\langle a^{\prime}\right|\mathbf{\hat{d}}\cdot\boldsymbol{\epsilon}_{\ell^{\prime}}\left|b\right\rangle \left\langle b\right|\mathbf{\hat{d}}\cdot\boldsymbol{\epsilon}_{\ell}\left|a\right\rangle }{\omega_{ba}-\omega-i\gamma_{ba}},\label{eq:amp}
\end{equation}

\noindent with $\omega$ ($\omega^{\prime}$) the frequency of the
incident (scattered) radiation, $\omega_{ba}$ ($\omega_{ba^{\prime}}$)
the resonant frequency between the initial state $\left|a\right\rangle $
(final state $\left|a^{\prime}\right\rangle $) and virtual state
$\left|b\right\rangle $, $\gamma_{ab}$ is, once again, the relaxation
coefficient for a transition between the $\left|b\right\rangle $
and $\left|a\right\rangle $ states (e.g., for a closed system it
equals half the Einstein spontaneous coefficient, $A_{ba}/2$, when
relaxation results only from spontaneous emission). As before, $n$
and $n^{\prime}$ are the number of photons in the initial and final
radiation states, while $\mathbf{\hat{d}}$ is the molecular electric
dipole moment operator and $\boldsymbol{\epsilon}_{\ell}$ the unit
vector associated to the linear polarization states with $\ell=\Vert$
or $\bot$. The quantities $T$ and $L^{3}$ are, respectively, the
period of interaction between the radiation and the molecule and the
fiducial volume of quantization for the radiation field \citep{Grynberg2010}.
From now on, we will assume that in equation (\ref{eq:amp}) the final
and initial linear polarization states are the same, as the difference
in the energy levels between states of differing polarization (i.e.,
the Zeeman splitting) will favor similar initial and final polarization
states (for a sufficiently long interaction period). This also implies
that $\omega_{ba^{\prime}}=\omega_{ba}$, $\omega=\omega^{\prime}$,
and $\left|a^{\prime}\right\rangle =\left|a\right\rangle $. Incidentally,
we recognize in equation (\ref{eq:amp}) the same type of second-order
interaction term encountered earlier in equation (\ref{eq:sigma_aa^2}). 

It follows that because we expect the relative phase shift to be very
small, i.e.,

\begin{equation}
\phi^{\prime}=\mathrm{Im}\left\{ S_{if,\Vert}-S_{if,\bot}\right\} \ll1,\label{eq:dS}
\end{equation}

\noindent we should resist the temptation to eliminate seemingly unimportant
differences. More precisely, if we write the frequency of the $\sigma$-transitions
as

\begin{equation}
\omega_{\pm}=\omega_{0}\pm\omega_{Z},\label{eq:w_sigma}
\end{equation}

\noindent with $\omega_{0}$ and $\omega_{Z}$ the $\pi$-transition
frequency and Zeeman splitting, respectively, then we should not approximate
$\omega_{\pm}\simeq\omega_{0}$ on the account that $\omega_{Z}\ll\omega_{0}$.
We require $\phi=N_{a}\phi^{\prime}\sim1$ for the linear-to-circular
polarization conversion effect to be measurable when the incident
radiation state is interacting with a large number $N_{a}$ of molecules.
Evidently the size of the volume of interaction is such that the relative
phase shift $\phi^{\prime}$ due to the scattering amplitude resulting
from the interaction with only one molecule, given by equation (\ref{eq:amp}),
can be extremely small while potentially still sufficient. Taking
this into account, we rewrite equation (\ref{eq:amp}) for the $\pi$-
and $\sigma$-transitions, respectively involving the $\left|n_{\Vert}\right\rangle $
and $\left|n_{\bot}\right\rangle $ states, with

\begin{eqnarray}
S_{if,\Vert} & = & -i\sin^{2}\left(\iota\right)\frac{T}{L^{3}}\frac{\sqrt{nn^{\prime}}}{2\epsilon_{0}\hbar}\left\Vert \left\langle b_{0}\right|\hat{d}_{\Vert}\left|a\right\rangle \right\Vert ^{2}\frac{\omega_{0}^{2}}{\omega\left(\omega_{0}-\omega-i\gamma_{b_{0}a}\right)}\label{eq:amp_para}\\
S_{if,\bot} & = & -i\sin^{2}\left(\iota\right)\frac{T}{L^{3}}\frac{\sqrt{nn^{\prime}}}{2\epsilon_{0}\hbar}\left\Vert \left\langle b_{\pm}\right|\hat{d}_{\bot}\left|a\right\rangle \right\Vert ^{2}\frac{\left(\omega_{0}\pm\omega_{Z}\right)^{2}}{\omega\left(\omega_{0}\pm\omega_{Z}-\omega-i\gamma_{b_{\pm}a}\right)},\label{eq:amp_perp}
\end{eqnarray}

\noindent where the summation on the virtual states was removed on
the account that one state $\left|b_{i}\right\rangle $ (with $i=0,\pm$)
will dominate independently for each transition because of the strong
resonance (note that $\omega\gg\omega_{Z}\gg\gamma_{b_{i}a}$ for
the transitions considered here). We have used this notation for the
virtual states $\left|b_{i}\right\rangle $ in equations (\ref{eq:amp_para})
and (\ref{eq:amp_perp}) to underline the different types of transitions
(i.e., $\pi$-transitions bring no change in magnetic quantum number
and obey $\Delta m_{J}=0$, while $\sigma$-transitions verify $\Delta m_{J}=\pm1$;
hence the notation). Also, the inclination angle of the magnetic field
relative to the scattering propagation direction (or the line of sight
to the observer) is given by $\iota$, from which $\mathbf{\hat{d}}\cdot\boldsymbol{\epsilon}_{\ell}=\hat{d}_{\ell}\sin\left(\iota\right)$
for the two states of linear polarization. In the numerical calculations
presented in the next section we will set $\iota=\pi/2$, effectively
setting the magnetic field in the plane of the sky.

If we now account for the population of molecules with which the radiation
interacts, then we must also consider the fact that their spectrum
(or velocity) will be spread over some normalized distribution function
$h\left(\omega\right)$. We can also substitute $T=l/c$ in the same
equations, where $l$ is the size of the region of interaction. We
then have at the frequency $\omega$ of the incident and scattered
photons

\begin{eqnarray}
\phi\left(\omega\right) & \simeq & -\sin^{2}\left(\iota\right)\frac{lN_{a}\sqrt{u\left(\omega\right)u^{\prime}\left(\omega\right)}}{2\epsilon_{0}c\hbar^{2}\omega^{2}}\left\{ \left\Vert \hat{d}_{\Vert,ba}\right\Vert ^{2}\int\frac{x^{2}\left(x-\omega\right)}{\left(x-\omega\right)^{2}+\gamma_{b_{0}a}^{2}}h\left(x\right)dx\right.\nonumber \\
 &  & \left.-\left\Vert \hat{d}_{\bot,ba}\right\Vert ^{2}\int\left[\frac{\left(x+\omega_{Z}\right)^{2}\left(x+\omega_{Z}-\omega\right)}{\left(x+\omega_{Z}-\omega\right)^{2}+\gamma_{b_{+}a}^{2}}+\frac{\left(x-\omega_{Z}\right)^{2}\left(x-\omega_{Z}-\omega\right)}{\left(x-\omega_{Z}-\omega\right)^{2}+\gamma_{b_{-}a}^{2}}\right]h\left(x\right)dx\right\} ,\label{eq:phi(w)}
\end{eqnarray}

\noindent where $N_{a}$ is the number of molecules in state $\left|a\right\rangle $
and $u\left(\omega\right)$ is the radiation energy density at frequency
$\omega$. For numerical calculations the terms $\left\Vert \hat{d}_{\ell,ba}\right\Vert ^{2}=\left\Vert \left\langle b_{i}\right|\hat{d}_{\ell}\left|a\right\rangle \right\Vert ^{2}$
($i=0,\pm$ and $\ell=\Vert,\bot$) can advantageously be related
to the corresponding Einstein spontaneous emission coefficient with
\citep{Cohen1988}

\begin{equation}
A_{\ell,ba}=\frac{\omega_{ba}^{3}\left\Vert \hat{d}_{\ell,ba}\right\Vert ^{2}}{3\pi\epsilon_{0}\hbar c^{3}}.\label{eq:A_ba}
\end{equation}

\noindent Notably, for $^{12}\mathrm{CO}$ we have $\left\Vert \hat{d}_{\Vert,ba}\right\Vert =\sqrt{2}\left\Vert \hat{d}_{\bot,ba}\right\Vert $
and $A_{b_{0}a}\simeq2A_{b_{\pm}a}$. It is therefore apparent from
equation (\ref{eq:phi(w)}) that any relative phase shift would vanish
had we approximated $\omega_{0}\pm\omega_{Z}\simeq\omega_{0}$.

\subsubsection{Circular Polarization of the $\mathit{^{12}CO\;\left(J=2\rightarrow1\right)}$
transition in Orion KL\label{sub:CO-OrionKL}}

Let us now provide an estimate of the importance of the effect for
the $^{12}\mathrm{CO}\;\left(J=2\rightarrow1\right)$ transition in
Orion KL. The region of interaction $l$ will be constrained by the
lifetime of the transition, which is determined by the relaxation
rate $\gamma_{ba}$, or the effective mean free path of a photon.
\citet{Plume2012} determined from observations of $\mathrm{CO}$
isotopologues that for Orion KL (in the Hot Core) the hydrogen number
density is $n_{\mathrm{H}_{2}}=10^{7}\;\mathrm{cm}^{-3}$ and the
temperature $T_{\mathrm{ex}}=150$ K, which together yield a collisional
quenching rate for $^{12}\mathrm{CO}$ of 

\begin{equation}
\gamma_{21}\simeq n_{\mathrm{H}_{2}}\left\langle \sigma v\right\rangle \approx1.3\times10^{-3}\:\mathrm{s}^{-1}.\label{eq:gamma21}
\end{equation}

\noindent This is more than three orders of magnitude greater than
$A_{21}^{\mathrm{CO}}=7\times10^{-7}\:\mathrm{s}^{-1}$ and we will
adopt this value for the corresponding relaxation rate, i.e, $\gamma_{ba}\equiv\gamma_{21}$
(at the specified temperature the momentum-rate transfer coefficient
$\left\langle \sigma v\right\rangle \approx10^{-10}\:\mathrm{cm^{3}\, s}^{-1}$;
see \citealt{Shull1990}). Accordingly, the associated region of interaction
equals $l_{\gamma}\approx c/\gamma_{21}\approx2.3\times10^{13}\:\mathrm{cm}$.
The effective mean path of a photon will depend on the absorption
coefficient $\alpha_{\omega}$ and the resonant scattering coefficient
$\sigma_{\omega}$ at the frequency $\omega$ of the photon. The absorption
coefficient \citep{Rybicki1979} is given by

\begin{equation}
\alpha_{\omega}=\frac{n_{\mathrm{CO}}g_{2}e^{-E_{1}/kT_{\mathrm{ex}}}}{Q_{\mathrm{CO}}\left(T_{\mathrm{ex}}\right)}\frac{\pi^{2}c^{2}A_{21}^{\mathrm{CO}}}{\omega^{2}}\frac{\hbar\omega}{kT_{\mathrm{ex}}}h\left(\omega\right)\label{eq:absortion}
\end{equation}

\noindent after integrating over the molecular population of the lower
state, with $g_{2}=5$ the degeneracy of the upper state, $E_{1}$
is the energy of the lower state ($E_{1}/k\simeq5.5\:\mathrm{K}$,
with $k$ the Boltzmann constant), $n_{\mathrm{CO}}\approx10^{3}\:\mathrm{cm}^{-3}$
the density of $^{12}\mathrm{CO}$ (i.e., a relative abundance of
approximately $10^{-4}$), and $Q_{\mathrm{CO}}\left(T_{\mathrm{ex}}\right)=54.6$
its partition function at $T_{\mathrm{ex}}=150$ K \citep{Pickett1998}. 

The resonant scattering coefficient is obtained through a similar
integration of the resonant scattering cross-section \citep{Grynberg2010}
over the molecular population of the lower state, which yields

\begin{equation}
\sigma_{\omega}=\frac{n_{\mathrm{CO}}g_{1}e^{-E_{1}/kT_{\mathrm{ex}}}}{Q_{\mathrm{CO}}\left(T_{\mathrm{ex}}\right)}\frac{3c^{2}}{\omega^{2}}4\pi^{3}\gamma_{21}h\left(\omega\right),\label{eq:scattering}
\end{equation}

\noindent with the degeneracy of the lower state $g_{1}=3$. If we
use a Gaussian distribution for the spectral profile at its maximum,
i.e., $h\left(\omega\right)=1/\left(\sqrt{2\pi}\Delta\omega\right)$,
with $\Delta\omega=9.4\times10^{7}\;\mathrm{rad\, s}^{-1}$ (i.e.,
a standard deviation of approximately $20\;\mathrm{km\, s}^{-1}$;
see below) and $\omega=1.4\times10^{12}\;\,\mathrm{rad\, s}^{-1}$
(i.e., $230.5\;\mathrm{GHz}$) we find that $\alpha_{\omega}=8.1\times10^{-17}\:\mathrm{cm}^{-1}$
and $\sigma_{\omega}=4.7\times10^{-11}\:\mathrm{cm}^{-1}$. The effective
mean path is then determined through \citep{Rybicki1979}

\begin{eqnarray}
l_{\mathrm{mp}} & = & \left[\alpha_{\omega}\left(\alpha_{\omega}+\sigma_{\omega}\right)\right]^{-1/2}\,\nonumber \\
 & \simeq & 1.6\times10^{13}\:\mathrm{cm}.\label{eq:meanfreepath}
\end{eqnarray}

\noindent We therefore find that both path lengths $l_{\gamma}$ and
$l_{\mathrm{mp}}$ are of similar sizes and little change would occur
whether we use one or the other. However, it is the case that $l_{\mathrm{mp}}\lesssim l_{\gamma}$
and we therefore set $l=l_{\mathrm{mp}}$ for the size of the region
of interaction in equation (\ref{eq:phi(w)}), which we rewrite here
for $^{12}\mathrm{CO}\;\left(J=2\rightarrow1\right)$ 

\begin{eqnarray}
\phi_{21}^{\mathrm{CO}}\left(\omega\right) & \simeq & -\sin^{2}\left(\iota\right)l_{\mathrm{mp}}^{4}\frac{n_{\mathrm{CO}}g_{1}e^{-E_{1}/kT_{\mathrm{ex}}}}{Q_{\mathrm{CO}}\left(T_{\mathrm{ex}}\right)}\frac{3\pi c^{2}A_{21}^{\mathrm{CO}}}{4\hbar\omega_{0}^{3}\omega^{2}}\sqrt{u\left(\omega\right)u^{\prime}\left(\omega\right)}\left\{ \int\frac{x^{2}\left(x-\omega\right)}{\left(x-\omega\right)^{2}+\gamma_{21}^{2}}h\left(x\right)dx\right.\nonumber \\
 &  & \left.-\frac{1}{2}\int\left[\frac{\left(x+\omega_{Z}\right)^{2}\left(x+\omega_{Z}-\omega\right)}{\left(x+\omega_{Z}-\omega\right)^{2}+\gamma_{21}^{2}}+\frac{\left(x-\omega_{Z}\right)^{2}\left(x-\omega_{Z}-\omega\right)}{\left(x-\omega_{Z}-\omega\right)^{2}+\gamma_{21}^{2}}\right]h\left(x\right)dx\right\} .\label{eq:phiCO}
\end{eqnarray}

\noindent In this equation the volume of the region of interaction
was set to $\approx l_{\mathrm{mp}}^{3}$. 

It is interesting to note that the integrals in equation (\ref{eq:phiCO})
can be combined and transformed such that 

\begin{equation}
\phi_{21}^{\mathrm{CO}}\left(\omega\right)\simeq\omega_{Z}^{2}\sin^{2}\left(\iota\right)l_{\mathrm{mp}}^{4}\frac{n_{\mathrm{CO}}g_{1}e^{-E_{1}/kT_{\mathrm{ex}}}}{Q_{\mathrm{CO}}\left(T_{\mathrm{ex}}\right)}\frac{3\pi c^{2}A_{21}^{\mathrm{CO}}}{4\hbar\omega_{0}^{3}\omega^{2}}\sqrt{u\left(\omega\right)u^{\prime}\left(\omega\right)}\, I\left(\omega\right),\label{eq:phiCO-2}
\end{equation}

\noindent where

\begin{eqnarray}
I\left(\omega\right) & = & \int\left\{ x^{2}\left(x-\omega\right)\left[3\left(x-\omega\right)^{2}-\gamma_{21}^{2}-\omega_{Z}^{2}\right]/\left[\left(x-\omega\right)^{2}+\gamma_{21}^{2}\right]\right.\nonumber \\
 &  & \left.+\left(x-\omega\right)\left(\omega^{2}-3x^{2}\right)+\gamma_{21}^{2}\left(3x-\omega\right)+\omega_{Z}^{2}\left(x+\omega\right)\right\} \frac{h\left(x\right)}{\Delta}dx,\label{eq:I(w)}
\end{eqnarray}

\noindent with $\Delta=\left[\left(x+\omega_{Z}-\omega\right)^{2}+\gamma_{21}^{2}\right]\left[\left(x-\omega_{Z}-\omega\right)^{2}+\gamma_{21}^{2}\right]$.
We thus find that the effect, or the relative phase shift $\phi_{21}^{\mathrm{CO}}$,
is proportional to the square of the magnitude of the plane of the
sky component of the magnetic field from the presence of the term
$\omega_{Z}^{2}\sin^{2}\left(\iota\right)$ in equation (\ref{eq:phiCO-2}).

\begin{figure}
\epsscale{0.6}\plotone{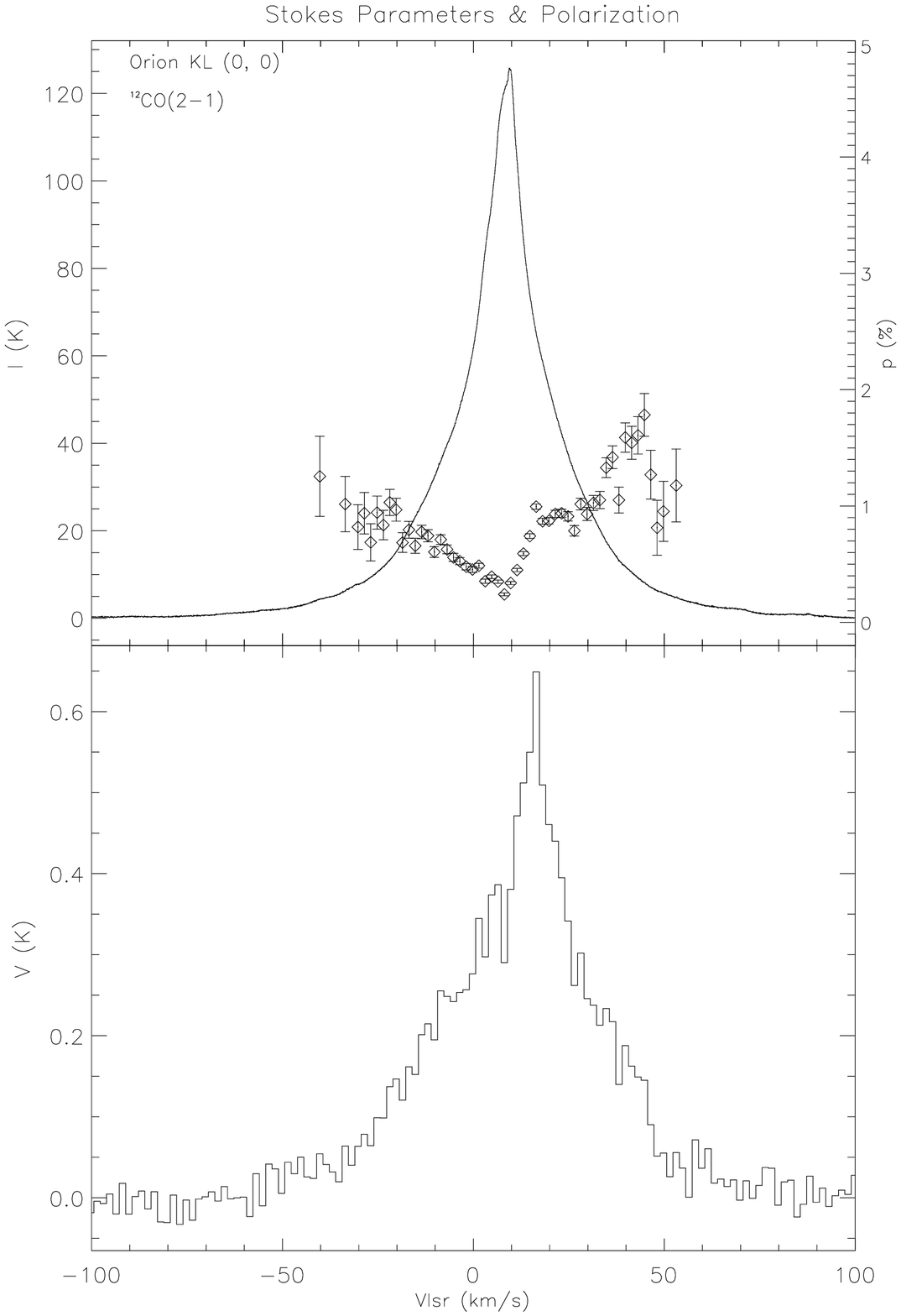}

\caption{\label{fig:CO_cp}Circular Polarization spectrum of the $^{12}\mathrm{CO}\;\left(J=2\rightarrow1\right)$
observations made at the peak position of Orion KL ($\mathrm{RA\,}(\mathrm{J2000})=05^{\mathrm{h}}35^{\mathrm{m}}14\fs5$,
$\mathrm{Dec\,}(\mathrm{J2000})=-05^{\circ}22\arcmin30\farcs4$) on
5 February 2012 at the CSO with FSPPol. \emph{Top:} Stokes $I$ spectrum,
uncorrected for telescope efficiency, and circular polarization levels
(symbols with uncertainty, using the scale on the right). All polarization
data satisfy $p\geq3\sigma_{p}$, where $p$ and $\sigma_{p}$ are
the polarization level and its uncertainty, respectively. \emph{Bottom:}
the Stokes $V$ spectrum, also uncorrected for telescope efficiency.
The frequency resolution of the Stokes $I$ spectrum is 61 kHz ($0.08\:\mathrm{km\, s^{-1}}$),
while the Stokes $V$ spectrum was smoothed by a factor of 20.}
\end{figure}

For the 10.4-m CSO telescope with an efficiency of approximately 60\%,
we can convert the antenna temperature to energy density according
to 

\begin{equation}
u\left(\omega\right)=5.42\times10^{-22}\,\frac{T_{\mathrm{A}}^{*}\gamma_{21}}{c}\approx2.3\times10^{-35}\, T_{\mathrm{A}}^{*}\qquad\mathrm{erg\, cm^{-3}}\label{eq:energy_density}
\end{equation}

\noindent while our circular and linear polarization spectra shown
in Figures \ref{fig:CO_cp} and \ref{fig:CO_lp} yield $u\left(\omega\right)\approx u^{\prime}\left(\omega\right)$
and $T_{\mathrm{A}}^{*}\lesssim0.2\:\mathrm{K}$ in the line wings
where polarization is detected. 

Using the parameter values already listed before, as well as $\iota=\pi/2$
and $\omega_{Z}\approx1.3\;\mathrm{rad\, s}^{-1}$ (i.e., $\approx0.2\:\mathrm{Hz}$,
with $B\approx1\:\mathrm{mG}$ \citep{Crutcher1999,Houde2009} and
$g_{J}^{\mathrm{CO}}\simeq-0.269$ \citep{Gordy1984}), we numerically
integrated equations (\ref{eq:phiCO-2}) and (\ref{eq:I(w)}) over
a Gaussian profile of the form

\begin{equation}
h\left(x\right)=\frac{1}{\sqrt{2\pi}\Delta\omega}e^{-\frac{1}{2}\left(\frac{x-\omega_{0}}{\Delta\omega}\right)^{2}}\label{eq:h(x)}
\end{equation}

\noindent chosen to approximately match the width of our observed
$^{12}\mathrm{CO}\;\left(J=2\rightarrow1\right)$ spectra shown in
Figures \ref{fig:CO_cp} and \ref{fig:CO_lp}. We also used a linear
radiation energy density profile $u\left(\omega\right)$ similar to
$h$ but with a peak temperature $T_{\mathrm{A}}^{*}=1\:\mathrm{K}$.
The result is shown in Figure \ref{fig:CO_phi_simu} where we find
that, for this set of parameters, the relative phase shift induced
by the resonant scattering process is potentially significant for
$^{12}\mathrm{CO}\;\left(J=2\rightarrow1\right)$ with a maximum value
for $\phi_{21}^{\mathrm{CO}}$ of approximately 30 rad. It is, however,
important to realize that these calculations are uncertain by a large
amount in view of the strong dependency of $\phi_{21}^{\mathrm{CO}}$
on some parameters. For example, a close look at equation (\ref{eq:phiCO-2})
reveals that $\phi_{21}^{\mathrm{CO}}$ varies inversely with the
fourth power of the gas density $n_{\mathrm{H}_{2}}$ and the third
power of the $^{12}\mathrm{CO}$ abundance; an increase of only a
factor of 2 for these parameters would bring the maximum value of
$\phi_{21}^{\mathrm{CO}}$ down to approximately 2 and 4 rad, respectively. 

The behavior of $\phi_{21}^{\mathrm{CO}}$ is different if the gas
density decreases to the point where the region of interaction is
defined by $l_{\gamma}$ instead of $l_{\mathrm{mp}}$ (see eqs. {[}\ref{eq:gamma21}{]}
to {[}\ref{eq:meanfreepath}{]} and the related discussion). We then
find that the relative phase shift varies inversely with the second
power of $n_{\mathrm{H}_{2}}$ and is proportional to the $^{12}\mathrm{CO}$
abundance (the same dependencies apply to other molecules with lower
abundances than $^{12}\mathrm{CO}$). Likewise, $\phi_{21}^{\mathrm{CO}}$
is also strongly dependent on the excitation temperature (mainly through
the partition function; with the inverse dependency of the molecular
abundance) and the magnetic field strength (proportional to its second
power through the Zeeman splitting). Whatever the case it is apparent
that the linear to circular polarization conversion effect is strongly
affected by even modest changes on a range of parameters, and within
their established uncertainties. For the present case, although there
are several combinations that would allow us to match the strength
of the polarization conversion effect found in our calculations to
the level of circular polarization we observe for $^{12}\mathrm{CO}\;\left(J=2\rightarrow1\right)$
in Orion KL, we will simply reduce the magnetic field strength from
$1\:\mathrm{mG}$ to $0.1\:\mathrm{mG}$ (resulting in a Zeeman splitting
$\omega_{Z}\simeq0.02\:\mathrm{Hz}$) and leave all other parameters
unchanged. Although this value for the magnetic field strength is
reasonable in general for star-forming regions at such gas densities,
there is no guarantee that it is right for this source. The corresponding
results can be found in Figure \ref{fig:CO_simu}, where the expected
relative phase shift $\phi_{21}^{\mathrm{CO}}$ for $^{12}\mathrm{CO}\;\left(J=2\rightarrow1\right)$
based on our resonant scattering model (solid curve; left vertical
scale), the underlying linear polarization radiation profile $u\left(v\right)$
(broken curve), and the corresponding Stokes $V$ spectrum (dot-broken
curve; right vertical scale) are shown. The Stokes $V$ spectrum is
proportional to $u\left(v\right)\sin\left(\phi_{21}^{\mathrm{CO}}\right)$,
and we set $2\alpha_{j}\beta_{j}=1$ (see eq. {[}\ref{eq:V}{]}).
Its maximum amplitude is now in line with that of our observations
shown in Figure \ref{fig:CO_cp}. We will discuss Figure \ref{fig:CO_simu}
in more details, especially its shape, in Section \ref{sec:Discussion}. 

The previous discussion raises the question as to what we should expect
observations to reveal in cases where $\phi_{21}^{\mathrm{CO}}$ becomes
very high. For example, measurements conducted with $^{12}\mathrm{CO}\;\left(J=2\rightarrow1\right)$
in molecular clouds will commonly probe gases of significantly lower
densities than is the case for Orion KL. Although changes in other
parameters (e.g., the excitation temperature and the magnetic field
strength) could somewhat offset the effect of a lower gas density,
it is likely that under such conditions the relative phase shift $\phi_{21}^{\mathrm{CO}}$
will reach significantly higher values (perhaps as much as several
hundred radians). The fact that the Stokes $V$ spectrum is proportional
to $\sin\left(\phi_{21}^{\mathrm{CO}}\right)$ will bring strong oscillations
as a function of the frequency (or velocity), which will then result
in a cancellation of any circular polarization across the spectrum
(given a finite spectral resolution). This cancellation will likely
be accentuated if several incident linear polarization modes are resonantly
scattered into the telescope beam. Finally, as a result of the Stokes
$u$ dependency on $\cos\left(\phi_{21}^{\mathrm{CO}}\right)$ (see
eq. {[}\ref{eq:U}{]}), only linear polarization parallel or perpendicular
to the magnetic field will be detected (i.e., Stokes $q$ in eq. {[}\ref{eq:Q}{]}).

\subsubsection{Circular Polarization of $\mathit{HCN}$ and $\mathit{HNCO}$ in
Orion KL\label{sub:HCN-HNCO}}

\begin{figure}
\epsscale{1.0}\plottwo{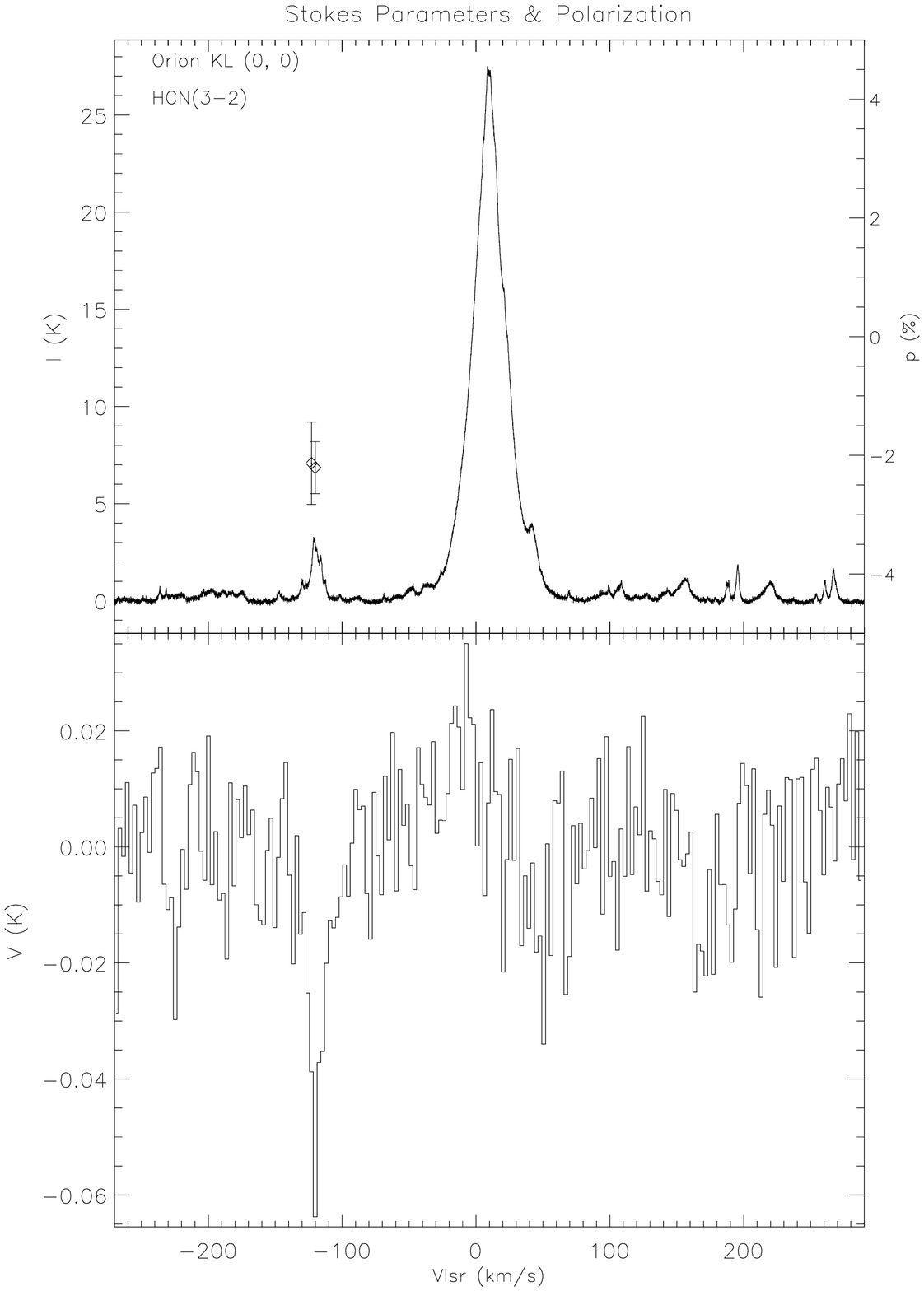}{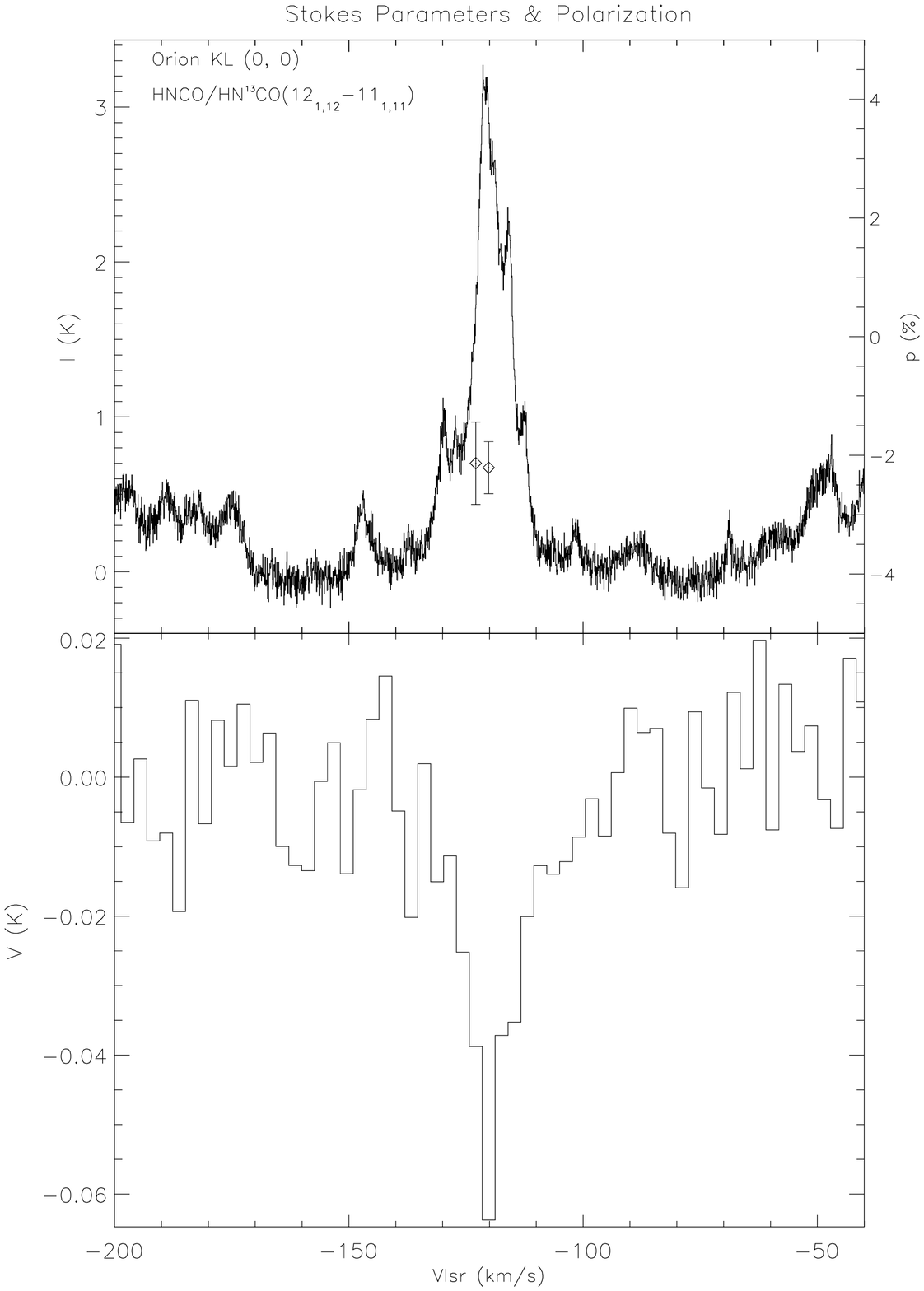}

\caption{\label{fig:HCN_cp}\emph{Left:} Same as Figure \ref{fig:CO_cp} but
for $\mathrm{HCN}\;\left(J=3\rightarrow2\right)$, obtained on 8 and
9 February 2012. The spectral feature located at $\approx-120\:\mathrm{km}\,\mathrm{s}^{-1}$
is a blend of lines from a few molecular species, most notably $\mathrm{HNCO}$
and $\mathrm{HN}^{13}\mathrm{CO}$ in the $\left(N_{K_{a}K_{c}}=12_{1,12}\rightarrow11_{1,11}\right)$
transitions; a close-up of this spectrum is shown on the right panel.
The frequency resolution of the Stokes $I$ spectra is 61 kHz ($0.07\:\mathrm{km\, s^{-1}}$),
while the Stokes $V$ spectra were smoothed by a factor of 40.}
\end{figure}

\begin{figure}
\epsscale{0.6}\plotone{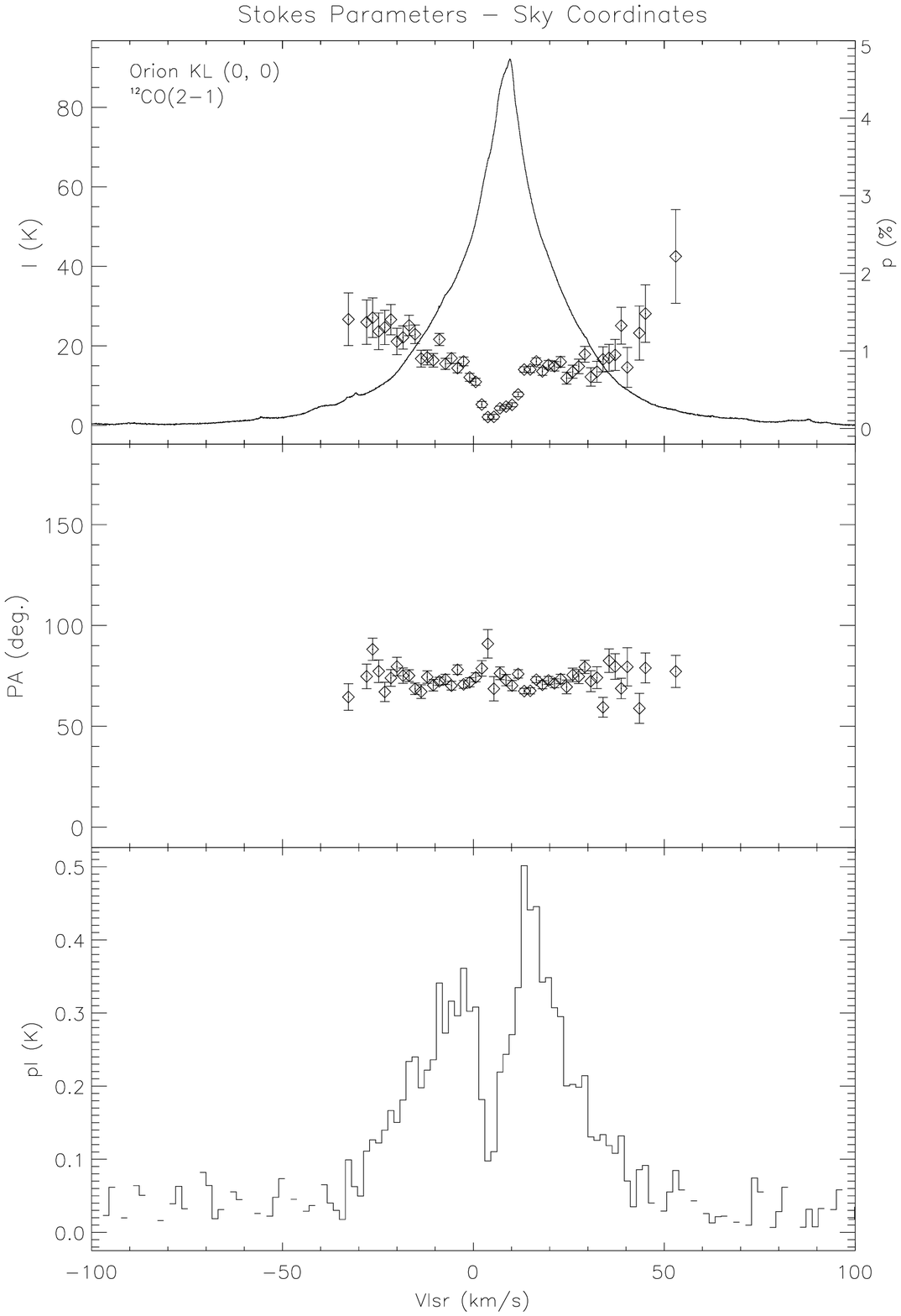}

\caption{\label{fig:CO_lp}Linear Polarization spectrum of the $^{12}\mathrm{CO}\;\left(J=2\rightarrow1\right)$
observations made at the peak position of Orion KL ($\mathrm{RA\,}(\mathrm{J2000})=05^{\mathrm{h}}35^{\mathrm{m}}14\fs5$,
$\mathrm{Dec\,}(\mathrm{J2000})=-05^{\circ}22\arcmin30\farcs4$) on
12 February 2012 at the CSO with FSPPol. \emph{Top:} Stokes $I$ spectrum,
uncorrected for telescope efficiency, and linear polarization levels
(symbols with uncertainty, using the scale on the right). All polarization
data satisfy $p\geq3\sigma_{p}$, where $p$ and $\sigma_{p}$ are
the polarization level and its uncertainty, respectively, and are
corrected for positive bias in the polarized flux. \emph{Middle: }Polarization
angle from north, increasing eastwards.\emph{ Bottom:} Polarized flux
$pI$, also uncorrected for telescope efficiency, but corrected for
positive bias due to noise. The frequency resolution of the Stokes
$I$ spectrum is 61 kHz ($0.08\;\mathrm{km\, s^{-1}}$), while the
polarized spectrum was smoothed by a factor of 20.}
\end{figure}

\begin{figure}
\epsscale{1.0}\plotone{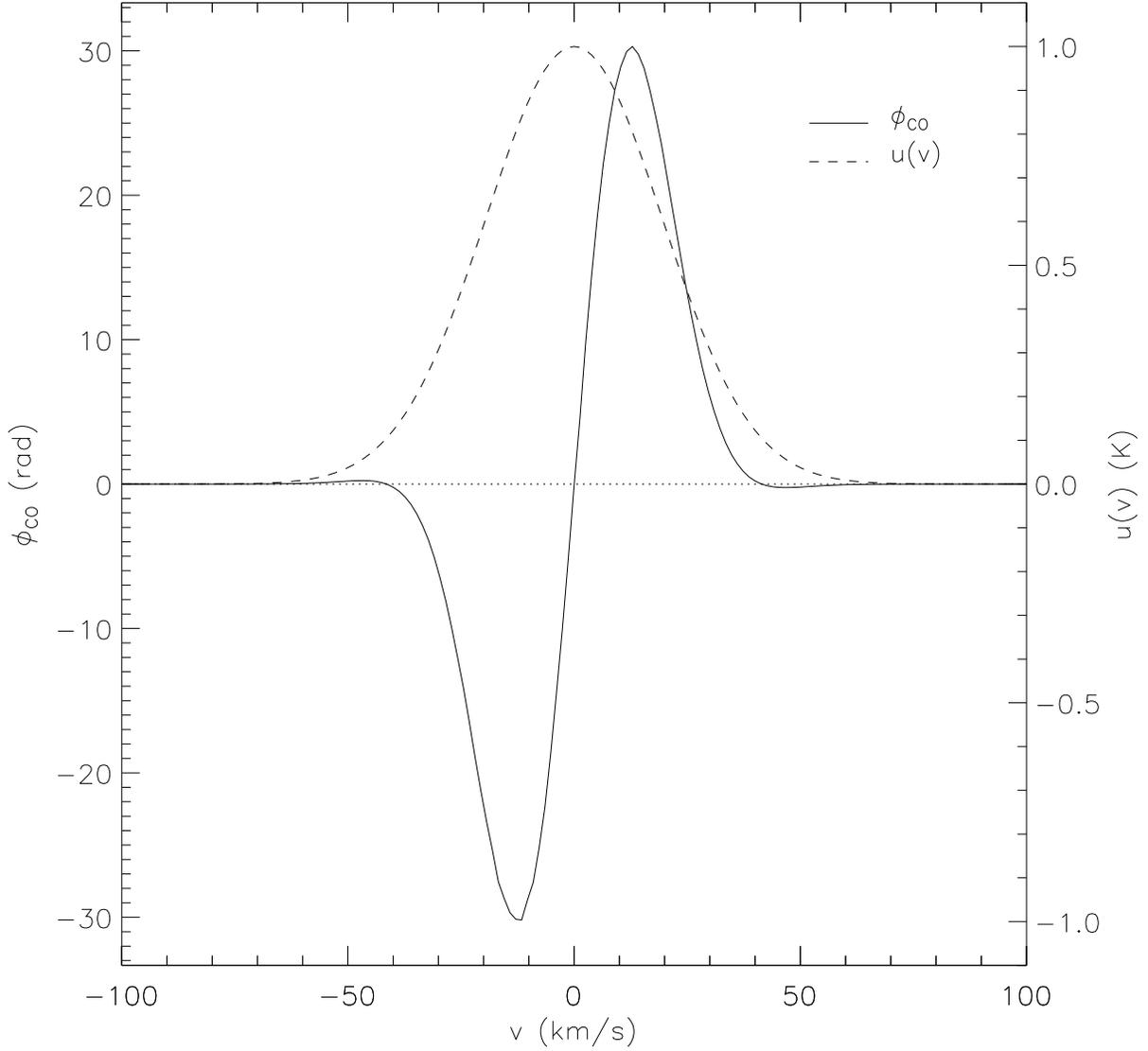}

\caption{\label{fig:CO_phi_simu}Numerical calculations of the expected relative
phase shift $\phi_{21}^{\mathrm{CO}}$ (solid curve) for $^{12}\mathrm{CO}\;\left(J=2\rightarrow1\right)$
in Orion KL based on our resonant scattering model; using the left
vertical scale. The magnetic field strength was set to 1 mG. The underlying
linear polarization radiation profile $u\left(v\right)$ is shown
with the broken curve, using the vertical scale on the right.}
\end{figure}

Evidently, our model applies equally to other linear molecules and
should therefore account for the lack of circular polarization in
our $\mathrm{HCN}\;\left(J=3\rightarrow2\right)$ spectrum of Figure
\ref{fig:HCN_cp}. For that molecule, its lower abundance, i.e., $\approx10^{-7}$,
implies that%
\footnote{Because of the larger value of the Einstein spontaneous emission coefficient
for this transition we set $\gamma_{\mathrm{HCN}}\simeq\gamma_{21}+A_{32}^{\mathrm{HCN}}/2$
for the relaxation rate \citep{Grynberg2010}; the value thus obtained
for $l_{\gamma}$ is slightly smaller than that for $^{12}\mathrm{CO}\;\left(J=2\rightarrow1\right)$
where the Einstein spontaneous emission coefficient could safely be
neglected.%
} $l=l_{\gamma}\approx1.7\times10^{13}\:\mathrm{cm}$ in equation (\ref{eq:phi(w)})
under similar conditions as for $^{12}\mathrm{CO}$ \citep{Blake1987,Schilke2001},
leaving the corresponding contribution to $\phi\left(\omega\right)$
basically unchanged. The reduction in molecular abundance is practically
cancelled out by an increase of approximately the same factor in the
Einstein spontaneous emission (i.e., $A_{32}^{\mathrm{HCN}}\simeq8\times10^{-4}\:\mathrm{s}^{-1}$)
in the numerator of the same equation. However, the Zeeman sensitivity
of HCN is approximately three times less than that of $^{12}\mathrm{CO}$
(i.e., $g_{J}^{\mathrm{HCN}}\simeq-0.0962$, from \citealt{Gordy1984}),
which implies a corresponding loss of about an order of magnitude
for $\phi\left(\omega\right)$. Taking all these factors into account,
as well as a decrease of about a factor of four for $u\left(\omega\right)$,
and the increases in $\omega_{0}$ and the partition function ($Q_{\mathrm{HCN}}\left(150\:\mathrm{K}\right)\simeq210$),
we find that the effect for $\mathrm{HCN}\;\left(J=3\rightarrow2\right)$
should be approximately two orders of magnitude weaker than for $^{12}\mathrm{CO}\;\left(J=2\rightarrow1\right)$,
again assuming similar physical conditions for the two molecular species.
This is consistent with our results of Figure \ref{fig:HCN_cp}. Moreover,
it is likely that the gas density probed by this molecular species
is higher given its significant critical density ($\approx10^{7}\;\mathrm{cm}^{-3}$).
An increase in $n_{\mathrm{H}_{2}}$ by only a factor of a few would
further reduce any circular polarization to even lower levels. 

The case of the $\mathrm{HNCO/HN}^{13}\mathrm{CO}\;\left(N_{K_{a}K_{c}}=12_{1,12}\rightarrow11_{1,11}\right)$
transitions is more difficult to analyze. These molecules are asymmetric
tops and will possess more complicated Zeeman spectra than linear
molecules like CO and HCN (we concentrate on HNCO in what follows).
We know, however, that the abundance of HNCO (i.e., $10^{-8}-10^{-9}$)
is less than CO by a factor of four to five orders of magnitude \citep{Tideswell2010}
and its Einstein spontaneous coefficient is comparable to $\mathrm{HCN}\;\left(J=3\rightarrow2\right)$
with $A^{\mathrm{HNCO}}=3\times10^{-4}\:\mathrm{s}^{-1}$ for these
transitions. However, the Zeeman sensitivity is likely to be significantly
higher as its electronic ground state possesses electronic spin. More
precisely, the upper and lower energy levels involved in the transitions
discussed here are part of triplet states with electronic spin $S=1$
\citep{Pickett1998}. The associated electronic spin contribution
to the Landé factor can be approximated with \citep{Gordy1984} 

\begin{equation}
g_{J}^{\mathrm{HNCO}}\simeq\frac{J\left(J+1\right)+S\left(S+1\right)-N\left(N+1\right)}{J\left(J+1\right)},\label{eq:g_J}
\end{equation}

\noindent which for the strongest of these lines, i.e., when $\Delta J=+1$
\citep{Pickett1998}, covers a range of $-0.182$ to $0.167$. The
Zeeman splitting between corresponding $\pi$- and $\sigma$-lines
can be approximated by multiplying these Landé factors by the Bohr
magneton. Taking into account the fact that the nuclear magneton is
used instead of the Bohr magneton for similar calculations with $^{12}\mathrm{CO}$
(and HCN), we expect the Zeeman splitting of the strongest $\mathrm{HNCO}$
$\left(N_{K_{a}K_{c}}=12_{1,12}\rightarrow11_{1,11}\right)$ lines
to be roughly a thousand times larger than for $^{12}\mathrm{CO}$.
This would lead to an approximately six orders of magnitude increase
in the value of $\omega_{Z}^{2}$ in equation (\ref{eq:phi(w)}).
Combining these changes (i.e., in molecular abundance, Einstein spontaneous
coefficient, and Zeeman sensitivity) with those for the partition
function ($Q_{\mathrm{HNCO}}\left(150\:\mathrm{K}\right)\simeq2800$,
about 50 times that of CO) and the radiation energy density (a reduction
of approximately 40 for similar levels of linear polarization), we
would expect the linear-to-circular polarization conversion effect
to be at approximately as strong for $\mathrm{HNCO}\;\left(N_{K_{a}K_{c}}=12_{1,12}\rightarrow11_{1,11}\right)$
as for $^{12}\mathrm{CO}\;\left(J=2\rightarrow1\right)$. Although
there is a significant level of uncertainty in the previous calculations,
this result is also consistent with our observations (Fig. \ref{fig:HCN_cp}).

\section{Discussion\label{sec:Discussion}}

The analysis presented in the previous section established that the
relative phase shift induced by the resonant scattering process between
linear polarization components parallel and perpendicular to the plane
of the sky component of the magnetic field can account for the levels
of circular polarization detected in the spectra presented in Figures
\ref{fig:CO_cp} and \ref{fig:HCN_cp}. It may be surprising that
a Zeeman splitting on the order of 0.1 Hz or less could be responsible
for such a significant effect across a spectral line that is on the
order of 10 MHz wide and centered at 230.5 GHz. But in the case of
$^{12}\mathrm{CO}\;\left(J=2\rightarrow1\right)$ the conversion of
linear to circular polarization is likely to be efficient in Orion
KL, although we again emphasize that small changes in some of the
main parameters (e.g., $\omega_{Z}$, $l_{\mathrm{mp}}$, or $n_{\mathrm{CO}}$)
can significantly affect our results. There is one aspect, however,
that requires further discussion: the expected profile of the circular
polarization Stokes $V$ spectrum.

\begin{figure}
\epsscale{1.0}\plotone{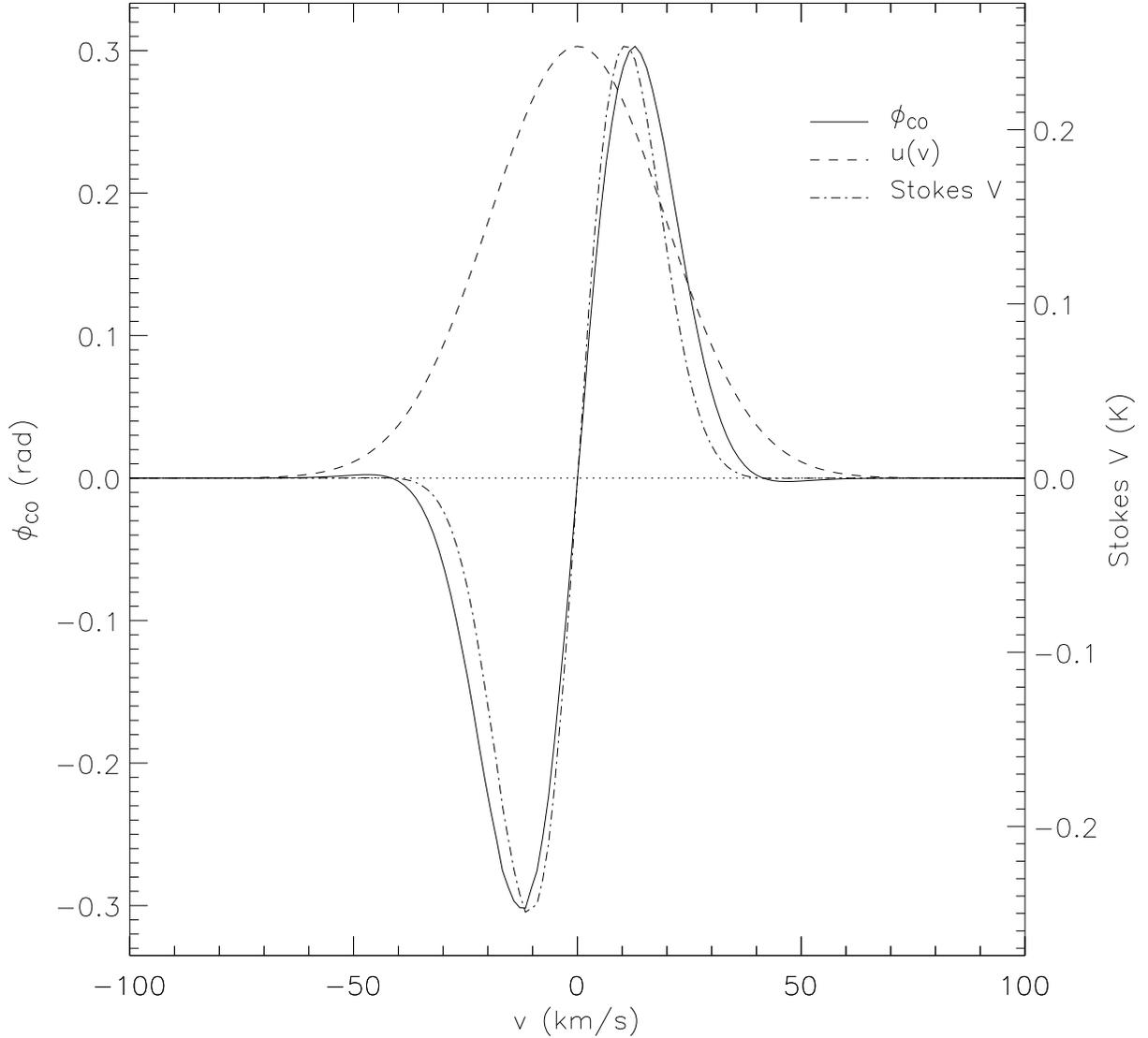}

\caption{\label{fig:CO_simu}Numerical calculations of the expected relative
phase shift $\phi_{21}^{\mathrm{CO}}$ (solid curve) for $^{12}\mathrm{CO}\;\left(J=2\rightarrow1\right)$
in Orion KL based on our resonant scattering model; using the left
vertical scale. The underlying linear polarization radiation profile
$u\left(v\right)$ is shown with the broken curve, has been normalized
in the figure but has a peak antenna temperature of 1 K. The corresponding
Stokes $V$ spectrum is also shown (dot-broken curve); using the vertical
scale on the right. This spectrum is proportional to $u\left(v\right)\sin\left(\phi_{21}^{\mathrm{CO}}\right)$.
For this simulation, and the other that follows, the magnetic field
strength was set to 0.1 mG.}
\end{figure}

\begin{figure}
\epsscale{1.0}\plotone{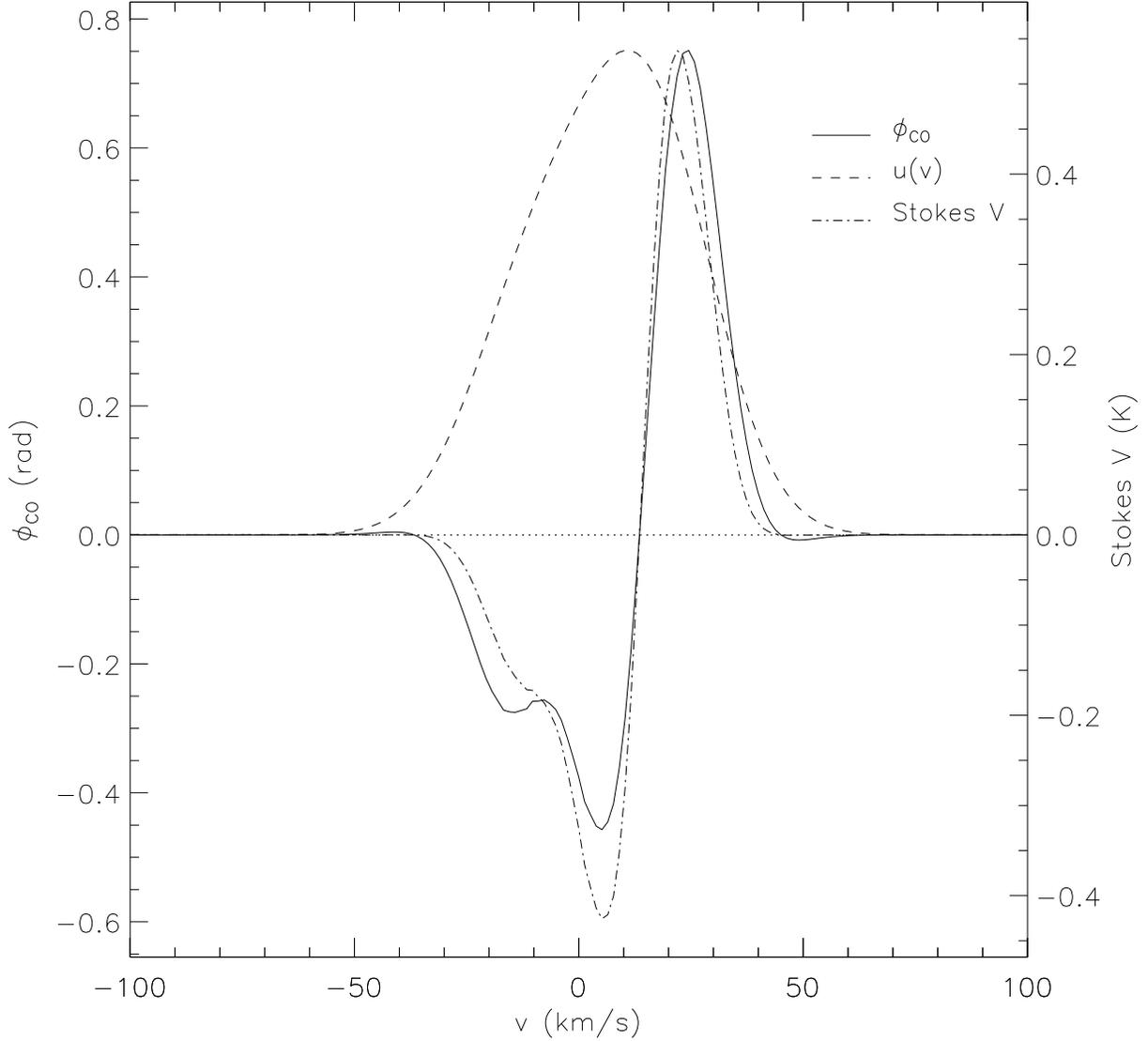}

\caption{\label{fig:CP_simu}Numerical calculations of the expected relative
phase shift $\phi_{21}^{\mathrm{CO}}$ (solid curve) for $^{12}\mathrm{CO}\;\left(J=2\rightarrow1\right)$
in Orion KL (using the left vertical scale) for a slightly uneven
underlying (normalized) linear polarization profile $u\left(v\right)$
of 1 K peak antenna temperature (broken curve). The corresponding
Stokes $V$ spectrum is also shown (dot-broken curve); using the vertical
scale on the right. Note the change in the resulting line profile
when compared with Figure \ref{fig:CO_simu}. The magnetic field strength
was set to 0.1 mG.}
\end{figure}

As was mentioned earlier, Figure \ref{fig:CO_simu} also shows the
Stokes $V$ profile that results from the numerical calculations of
$\phi_{21}^{\mathrm{CO}}$ discussed in Section \ref{sub:CO-OrionKL}
(the dot-broken curve; using the vertical scale in the right side
of the graph). We approximated this spectrum with the following function
(in units of Kelvin)

\begin{equation}
V\left(\omega\right)=e^{-\frac{1}{2}\left(\frac{\omega-\omega_{0}}{\Delta\omega}\right)^{2}}\sin\left[\phi_{21}^{\mathrm{CO}}\left(\omega\right)\right],\label{eq:Stokes V}
\end{equation}

\noindent i.e., we set $2\alpha_{j}\beta_{j}\approx1$ (see eqs. {[}\ref{eq:V}{]}
and {[}\ref{eq:psi_j_again}{]}) and, as is apparent in equation (\ref{eq:Stokes V}),
used a Gaussian profile for $u\left(\omega\right)$ with a peak antenna
temperature of 1 K for the radiation associated with the $\left|n_{\Vert}\right\rangle $
and $\left|n_{\bot}\right\rangle $ incident states. We also converted
the abscissa to a velocity scale to ease the comparison with the observed
spectrum of Figure \ref{fig:CO_cp}. The most obvious discrepancy
between the two spectra is the fact that our calculations yield an
antisymmetric profile, while the observations do not. In fact, the
calculated profile is not unlike the typical Stokes $V$ spectrum
expected from Zeeman sensitive molecular species and transitions.
A detailed comparison between calculations and observations is complicated
by the fact that our model contains several approximations. For example,
the Gaussian profile chosen for our calculations does not perfectly
match that of the measured linear polarization spectrum of Figure
\ref{fig:CO_lp}, and the core of the observed polarization line profiles
(linear and circular) are very likely dominated by instrumental polarization,
which our model does not consider. But it is important to realize
that several factors can affect the shape of the calculated Stokes
$V$ spectrum: 
\begin{enumerate}
\item The perfect antisymmetry seen in the results of Figure \ref{fig:CO_simu}
stems from the symmetry of the underlying Gaussian used for the calculations.
This is made clearer in Figure \ref{fig:CP_simu} where another calculation
for the same parameters used for Figure \ref{fig:CO_simu} is shown,
but with the Gaussian profile of equations (\ref{eq:h(x)}) and (\ref{eq:Stokes V})
replaced with a slightly uneven line shape. The result is a circular
polarization spectrum that is markedly broader on one side (where
$v\lesssim15\;\mathrm{km\, s^{-1}}$) than the other. The same would
happen for a typical Stokes $V$ spectrum from a Zeeman sensitive
transition for a slightly uneven Stokes $I$ profile, since they are
linked through a derivative. But it is important to note that, in
our case, it is the line profile of the incident (background) linear
polarization radiation that is in question, which may be different
than the Stokes $I$ spectrum. 
\item In the calculations of the integral contained in equation (\ref{eq:phi(w)})
(or {[}\ref{eq:I(w)}{]}) we assumed that a photon at a frequency
$\omega$ will scatter off all molecules at any other frequencies
covered under the profile $h\left(\omega\right)$. We must realize
that this is not likely to be the case since the size of the molecule-radiation
region of interaction is relatively small (i.e., $\sim10^{13}$ cm
or $\sim10^{-5}$ pc; see eq. {[}\ref{eq:meanfreepath}{]}) whereas
molecules belonging to different velocity ranges are likely to be
separated by larger distances within Orion KL (the $^{12}\mathrm{CO}\;\left(J=2\rightarrow1\right)$
spectrum covers a velocity range of $\approx\pm50\:\mathrm{km\, s^{-1}}$).
This could significantly affect the shape of the resulting Stokes
$V$ spectrum.
\item The numerical calculations that produced Figure \ref{fig:CO_simu}
considered a single incident radiation mode. But we know from the
related discussion in Section \ref{sub:simple} (see eqs. {[}\ref{eq:total_state}{]}
to {[}\ref{eq:D}{]}) that even at a single frequency several incident
modes, coming from different orientations with potentially differing
linear polarization states, take part in the resonant scattering process.
This brings an averaging process that will certainly affect the line
shape of the circular polarization profile.
\item There also exists an averaging process due to the size of the telescope
beam. At the distance of Orion KL (i.e., approximately 450 pc) our
telescope beam of $\approx32\arcsec$ covers a region of approximately
0.07 pc or $10^{17}$ cm, which is several orders of magnitude larger
than the region of interaction. Our observations therefore contain
contributions from several radiation-molecule interaction regions,
where key parameters that affect the Stokes $V$ line profile are
likely to vary.
\end{enumerate}
Although these considerations reveal the complexity of the problem
and that we perhaps should not expect our observations to closely
match the profile resulting from our simplified model and presented
in Figure \ref{fig:CO_simu}, it is nevertheless possible to find
relatively simple conditions that would allow us to calculate Stokes
$V$ profiles that are consistent with our circular polarization detection
in $^{12}\mathrm{CO}\;\left(J=2\rightarrow1\right)$.

\begin{figure}
\epsscale{1.0}\plotone{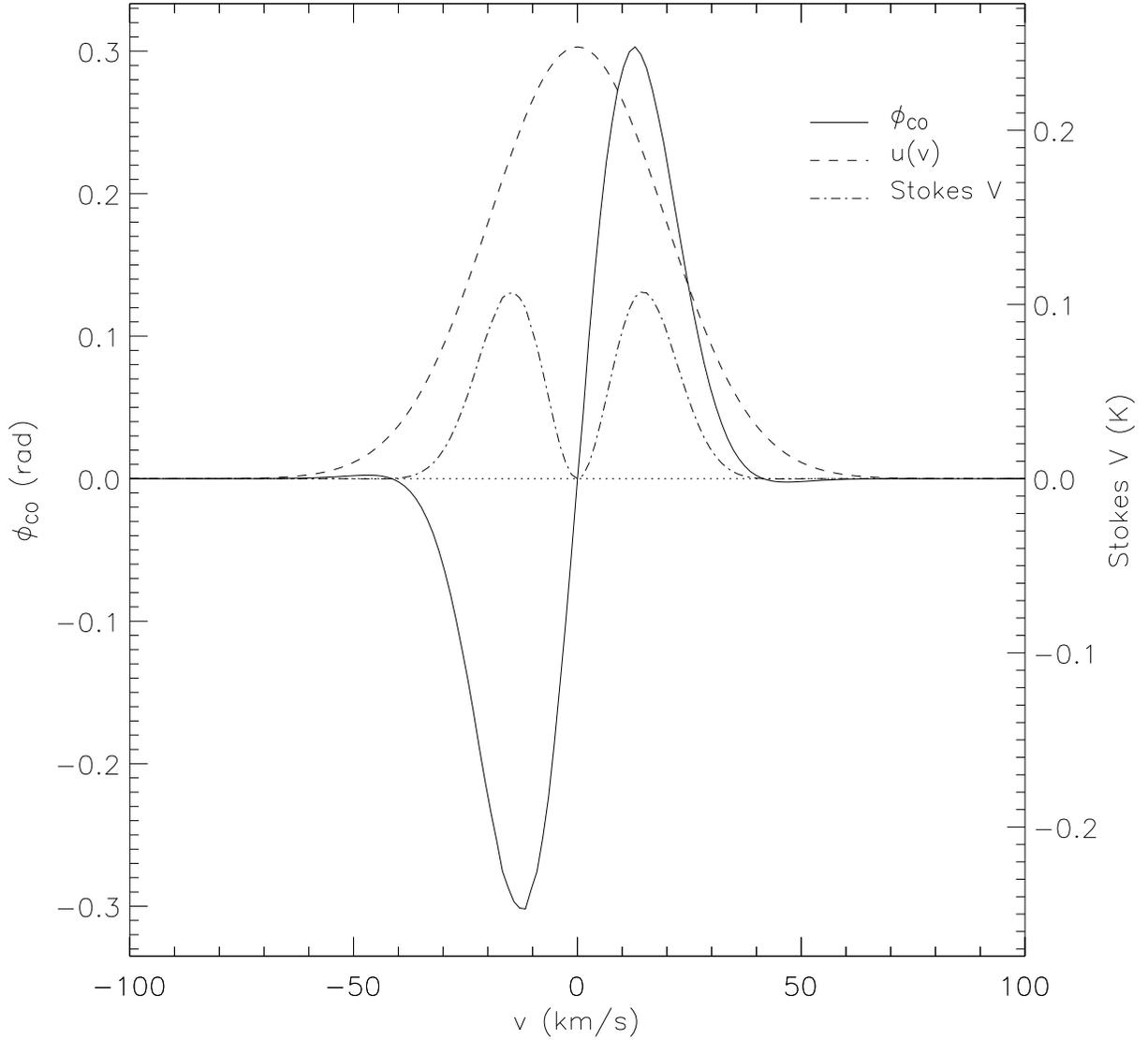}

\caption{\label{fig:CP_sym}Numerical calculations of the expected relative
phase shift $\phi_{21}^{\mathrm{CO}}$ (solid curve) for $^{12}\mathrm{CO}\;\left(J=2\rightarrow1\right)$
in Orion KL (using the left vertical scale) for a case where the angle
$\theta$ between the incident linear polarization and the magnetic
field orientation varies linearly with frequency (or velocity) at
a rate of $d\theta/dv=-1\:\mathrm{deg/(km\, s^{-1})}$ (with $\theta=0$
at $v=0$). The underlying linear polarization radiation profile $u\left(v\right)$,
shown with the broken curve, has been normalized in the figure but
has a peak antenna temperature of 1 K. The corresponding Stokes $V$
spectrum is also shown (dot-broken curve); using the vertical scale
on the right. This spectrum is proportional to $-2\sin\left(\theta\right)\cos\left(\theta\right)u\left(v\right)\sin\left(\phi_{21}^{\mathrm{CO}}\right)$
(see eq. {[}\ref{eq:V}{]}). Note that the Stokes $V$ spectrum now
has a symmetric profile about $v=0$. The magnetic field strength
was set to 0.1 mG.}
\end{figure}

To do so we consider incident radiation modes for which $\theta\left(v\right)$
varies with velocity (frequency) and possesses some distribution about
the orientation of the foreground magnetic field on the plane of the
sky. Since the linear to circular conversion of polarization will
only occur when $\theta\left(v\right)\neq0$ (and $\pi/2$) it is
possible that most linear polarization signals at frequencies satisfying
this condition will be efficiently converted to circular polarization
leaving only a well defined orientation angle (i.e., $\theta=0$)
for the outgoing linear polarization radiation. Our linear polarization
spectrum of Figure \ref{fig:CO_lp} indeed shows a well defined polarization
angle across the spectral line. If the distribution of $\theta\left(v\right)$
is not uniform about $\theta=0$ and, for example, is positive in
some regions and negative in others, then the shape of the circular
polarization spectrum would clearly depart from the asymmetric profile
displayed in Figure \ref{fig:CO_simu} (or even \ref{fig:CP_simu}).
This is exemplified in Figure \ref{fig:CP_sym} where have computed
the Stokes $V$ spectrum for conditions similar to those used for
Figure \ref{fig:CO_simu}, with the exception that the angle $\theta\left(v\right)$
between the incident linear polarization and the magnetic field orientation
varies linearly with velocity at a rate of $d\theta/dv=-1\:\mathrm{deg/(km\, s^{-1})}$
(with $\theta=0$ at $v=0$). The resulting Stokes $V$ spectrum follows
from equations (\ref{eq:V}) and (\ref{eq:phiCO-2}) (and eq. {[}\ref{eq:Stokes V}{]}).
It is then found that the presence of the factor $\alpha_{j}=\sin\left(\theta\right)$
in this equation results in a symmetric profile about $v=0$ that
is consistent with our circular polarization measurements on $^{12}\mathrm{CO}\;\left(J=2\rightarrow1\right)$.
As stated above, the core of this spectral line is very likely dominated
by instrumental polarization and the true polarization level there
should be zero or close to zero, as in our Figure \ref{fig:CP_sym}.
Although this does not ensure that our model can perfectly account
for our observations across the whole spectrum, the results of Figure
\ref{fig:CP_sym} are consistent with our measurements away from the
core of the spectral line where signals in the Stokes $V$ profile
have the same sign.

We also note that the levels of circular polarization predicted by
our model are likely to be consistent with the IK Tau SiO $v=1$,
$v=2$, $\left(J=1\rightarrow0\right)$ (at 43.1 GHz and 42.8 GHz,
respectively) observations of \citet{Cotton2011}. Although a more
precise analysis would be necessary to ensure this (we are not, for
example, accounting for possible maser saturation), we can make an
approximate assessment of the strength of the effect for SiO relative
to CO using our previous calculations. If we take the SiO $v=2$,
$\left(J=1\rightarrow0\right)$, 42.8 GHz transition as an example,
we know that although the abundance is one to two orders of magnitude
less than $^{12}\mathrm{CO}$ \citep{Decin2012}, its Einstein spontaneous
coefficient is about an order of magnitude greater at $A\simeq4\times10^{-6}\:\mathrm{s^{-1}}$,
while its Zeeman sensitivity is only a factor of two or so lower (\citet{Davis1974}
find $g_{J}^{\mathrm{SiO}}\simeq-0.154$ to $-0.155$ for $v=0$,
1, and 2). If we add to this the fact that the frequency of SiO $v=2$,
$\left(J=1\rightarrow0\right)$ is less than five times that of $^{12}\mathrm{CO}\;\left(J=2\rightarrow1\right)$
and set the spectral line width to match their observations (on the
order of $1\;\mathrm{km\: s}^{-1}$), then we find that the polarization
conversion effect is likely to be important for SiO for a significant
range of gas densities and magnetic field strengths. A similar result
is expected for the SiO $v=1$, $\left(J=1\rightarrow0\right)$, 43.1
GHz transition. These calculations could then resolve the known problem
of high circular polarization levels found in maser transitions that
require unreasonably large magnetic field strengths when interpreted
within the context of the Zeeman effect \citep{Watson2009}. Furthermore,
our model is also consistent with other observations of SiO maser
lines in evolved stars that showed a correlation between measured
levels of linear and circular polarizations, no correlation between
levels of circular polarization and Stokes $I$, as well as a case
where circular polarization is more important than linear polarization
\citep{Herpin2006}. 

We stress, however, that although some SiO maser Stokes $V$ line
profiles presented in \citet{Cotton2011} match well the asymmetric
profile resulting from our simplest calculations shown in Figures
\ref{fig:CO_simu} and \ref{fig:CP_simu} (i.e., when the angle $\theta$
between the incident linear polarization and the magnetic field orientation
is constant across the spectrum), many of their spectra show (quasi-)
symmetric Stokes $V$ profiles that are different from anything we
presented. More precisely, if the $^{12}\mathrm{CO}\;\left(J=2\rightarrow1\right)$
line in Orion KL is expected to yield little to no polarization at
or near the center of the line, the same cannot apply to maser lines
in view of the nature of the stimulated emission process. That is,
several of \citet{Cotton2011} Stokes $V$ spectra show a maximum
(in the absolute sense) near the center of the line, not a near-zero
value. It follows that the symmetric Stokes $V$ spectrum of Figure
\ref{fig:CP_sym} cannot, by itself, explain the aforementioned results
of \citet{Cotton2011}. It is still possible, however, that the presence
of several maser spots within the telescope beam could lead to such
(quasi-) symmetric line profiles under the assumption that the underlying
Stokes $V$ spectrum of a given maser is as shown in Figure \ref{fig:CP_sym}
and the masers' systemic velocities are spread across the extent of
the observed spectral line. The combination of these relatively-shifted
Stokes $V$ spectra (i.e., as in Fig. \ref{fig:CP_sym}) could thus
yield something akin to the (quasi-) symmetric line profile of \citet{Cotton2011}.
However, it is perhaps safer at this point to await the results of
an analysis similar to the one presented in this paper, but specifically
tailored to maser emission, to find out if our resonant scattering
model can account for all aspects of SiO maser circular polarization
spectra with a minimum number of assumptions. This we intend to attempt
in a future publication. 

Our results also have implications for Zeeman measurements in non
masing environments (e.g., in molecular clouds). For example, it is
clear from our discussion of Section \ref{sub:HCN-HNCO} that HNCO
is a Zeeman sensitive molecule (in contrast to CO and HCN). However,
it is likely that our observations of Figure \ref{fig:HCN_cp} would
be dismissed as due to instrumental artifacts when analyzed within
the context of the Zeeman effect (e.g., leakage from Stokes $I$ to
Stokes $V$). But when studied in conjunction with the HCN result
presented in the same spectrum and that for $^{12}\mathrm{CO}\;\left(J=2\rightarrow1\right)$
shown in Figure \ref{fig:CO_cp}, another interpretation is warranted.

\section{Conclusion\label{sec:Conclusion}}

We presented measurements of circular polarization from rotational
spectral lines of molecular species in Orion KL obtained at the Caltech
Submillimeter Observatory with the Four-Stokes-Parameter Spectra Line
Polarimeter. We measured levels of polarization of up to 1 to 2\%
for $^{12}\mathrm{CO}\;\left(J=2\rightarrow1\right)$ and $\mathrm{HNCO/HN}^{13}\mathrm{CO}\;\left(N_{K_{a}K_{c}}=12_{1,12}\rightarrow11_{1,11}\right)$,
while none was detected for $\mathrm{HCN}\;\left(J=3\rightarrow2\right).$
We further presented a physical model based on resonant scattering
in an attempt to explain our observations, through the conversion
of linear polarization to circular polarization. We also showed that
this effect is proportional to the square of the magnitude of the
plane of the sky component of the magnetic field, and therefore opens
up the possibility of measuring this parameter from circular polarization
measurements of Zeeman insensitive molecules. We intend to study this
in an upcoming paper.

\acknowledgements{M.H. is grateful to C. Cohen-Tannoudji of the Ecole Normale Supérieure
de Paris for an insightful discussion. M.H.'s research is funded through
the NSERC Discovery Grant, Canada Research Chair, and Western's Academic
Development Fund programs. T. H. is funded by the Alexander von Humboldt
foundation in Germany. The Caltech Submillimeter Observatory is operated
by the California Institute of Technology under cooperative agreement
with the National Science Foundation (AST-0838261). }



\end{document}